\pdfoutput=1 \pdfsuppresswarningpagegroup=1 

\documentclass[%
  journal=nalefd,%
  manuscript=letter,%
  layout=twocolumn,%
]{achemso}

\makeatletter
\let\l@addto@macro\relax
\makeatother
\usepackage[fontsize=10pt]{scrextend}
\usepackage{times}
\usepackage{pdfpages}
\makeatletter
\AtBeginDocument{\let\LS@rot\@undefined}
\makeatother
\usepackage[nooneline]{subfigure}
\usepackage{amsfonts,amsmath,amssymb,mathtools,bm}
\usepackage{soul}
\usepackage{graphicx}

\graphicspath{ {./Figs/} }

\newcommand{\Fref}[1]{Figure~\ref{#1}}
\newcommand{\Fsref}[1]{Figures~\ref{#1}}
\newcommand{\Eqref}[1]{eq.~(\ref{#1})}

\renewcommand{\eqref}[1]{(\ref{#1})}

\newcommand{\SI}{Supporting Information}
\newcommand{\ep}{\varepsilon}

\newcommand{\epm}{{\ep_\text{m}}}
\newcommand{\Weff}{{W_\text{eff}}}
\newcommand{\Gup}{{G_\text{up}}}
\newcommand{\Gp}{{G_\text{p}}}
\newcommand{\mb}{\boldsymbol}

\DeclareMathOperator{\area}{area}
\DeclareMathOperator{\Tr}{Tr}
\newcommand{\unit}[1]{\,\textup{#1}}

\author{Gareth W. Jones}
\affiliation{School of Mathematics, The University of Manchester, Manchester,
M13 9PL, England}
\altaffiliation{These authors contributed equally to this work.}

\author{Dario Andres Bahamon}
\affiliation{MackGraphe\,--\,Graphene and Nano-Materials Research Center, 
Mackenzie Presbyterian University, Rua da Consola\c{c}\~{a}o 896, 01302-907, 
S\~{a}o Paulo, SP, Brazil}
\altaffiliation{These authors contributed equally to this work.}

\author{Antonio H. Castro Neto}
\affiliation{Department of Physics, National University of Singapore,
2 Science Drive 3, Singapore 117542}
\alsoaffiliation{Centre for Advanced 2D Materials, National University of 
Singapore, 6 Science Drive 2, Singapore 117546}

\author{Vitor~M.~Pereira}
\email{vpereira@nus.edu.sg}
\phone{+65 6601 3642}
\affiliation{Department of Physics, National University of Singapore,
2 Science Drive 3, Singapore 117542}
\alsoaffiliation{Centre for Advanced 2D Materials, National University of 
Singapore, 6 Science Drive 2, Singapore 117546}

\title{Quantized transport, strain-induced perfectly conducting modes and valley 
filtering on shape-optimized graphene Corbino devices\footnote{This work has 
been published in Nano Letters on 4 August 2017, with the DOI 
10.1021/acs.nanolett.7b01663.}}

\begin{document}

\newpage

\begin{abstract}
The extreme mechanical resilience of graphene 
\cite{Booth2008Macroscopic,Lee2008Measurement} and the peculiar coupling it 
hosts between lattice and electronic degrees of freedom 
\cite{Kane1997Size,Suzuura2002Phonons,Vozmediano2010Gauge} have spawned a strong 
impetus towards strain-engineered graphene where, on the one hand, strain 
augments the richness of its phenomenology and makes possible new concepts for 
electronic devices \cite{Pereira2009Strain,Tomori2011Introducing} and, on the 
other hand, new and extreme physics might take place 
\cite{GuineaNP2010,SanJose2011Rippling,Low2011Gaps,Wu2011ValleyDependent, 
Klimov2012Electromechanical,Lu2012Transforming,Juan2012SpaceDependent, 
Abanin2012Interaction}. 
Here, we demonstrate that the shape of substrates supporting graphene 
sheets can be optimized for approachable experiments where strain-induced 
pseudomagnetic fields (PMF) can be tailored by pressure for directionally 
selective electronic transmission and pinching-off of current flow down to the 
quantum channel limit.
The Corbino-type layout explored here furthermore allows filtering of charge 
carriers according to valley and current direction, which can be used to inject 
or collect valley-polarized currents, thus realizing one of the basic elements 
required for valleytronics.
Our results are based on a framework developed to realistically determine the 
combination of strain, external parameters, and geometry optimally compatible 
with the target spatial profile of a desired physical property --- the 
PMF in this case. Characteristic conductance profiles are analyzed
through quantum transport calculations on large graphene devices having the 
optimal shape.
\end{abstract}

\newpage

\section{Introduction}%
An extremely active and promising area of research in strain-engineered 
two-dimensional electronic systems is the realization of graphene-based 
nanostructures with tailored pseudomagnetic fields (PMF). The PMF is a 
fruitful concept in strained graphene arising from the fact that, in the 
vicinity of the Dirac point, the description of the coupling between lattice 
deformations and electrons can be captured by two types of strain-induced 
fields: a displacement potential that couples to electrons as an electrostatic 
potential, and a strain-induced gauge field that couples similarly to a magnetic 
field \cite{Suzuura2002Phonons}. Since the former are expected to be effectively 
screened by the charge carriers, electrons are then expected to respond to 
strain mostly through this pseudomagnetic effect.
Since (pseudo)magnetic fields are very effective means of guiding the motion and 
confining charged carriers, this naturally led to proposals and studies of many 
analogues of magnetic devices using graphene without actual magnetic fields, 
such as pseudomagnetic barriers 
\cite{Pereira2009Strain,Pellegrino2011Transport,Bahamon2013ConductanceAcross, 
Yesilyurt2016PerfectValley} or pseudomagnetic quantum dots 
\cite{GuineaNP2010,Qi:2013}. 

There are significant advantages of using pseudomagnetic fields in graphene 
electronic devices at the nanoscale. 
The first is that electrons in graphene are not easily confined by electric 
fields due to the Klein tunneling effect \cite{Katsnelson2006Chiral} arising 
from the relativistic-like behavior of electrons in this system. This has 
been a perennial difficulty and disadvantage of graphene-based electronics, 
while PMFs can efficiently localize electrons or act as tunneling barriers 
\cite{Pereira2009Strain}.
The second advantage is that PMFs are only felt by the electrons in graphene, 
and they can be confined to regions as narrow as 5\,--\,10\,nm 
\cite{LevySci2010,Klimov2012Electromechanical}. This is unlike real magnetic 
fields which are felt not only by the target electronic system but also by 
whole the environment nearby. In addition, fabricating real magnetic barriers of 
nanoscale size is practically impossible and their magnetic field would also be 
fixed in principle, not reconfigurable or tuneable.
Finally, the magnitude of PMFs that can be achieved in graphene is remarkably 
high and in excess of 300\,T 
\cite{LevySci2010,Lu2012Transforming,Klimov2012Electromechanical} which means 
that values of the order of tens of Tesla are easily obtainable, and strong 
enough for many electronic applications.

There is a key challenge that has persisted since the early proposals for 
strain-engineered graphene: the ability to tailor the spatial 
distribution of the PMF in graphene hinges on the ability to define specific 
non-uniform strain profiles, which is a difficult experimental prospect, 
especially at the micro and nanoscale. Moreover, whereas it is a 
straightforward theoretical task to start from a given strain field and 
determine the associated PMF and its impact in the electronic and transport 
properties, it is much less trivial and non-unique to request the opposite. 
Yet, it is precisely the opposite (i.e., specifying the PMF patterns in real 
space and obtaining the necessary strain profiles) that is of most direct interest 
for application in electronic devices because, in nearly any such case, one 
should be able to specify how, where, and to which extent we need electrons to 
be confined, which can be easily done reasoning in terms of PMF patterns alone, 
but is generally hard to anticipate in terms of strain alone. The central 
theoretical question for actual realizations and applications of 
strain-engineered graphene then becomes: how should one design the 
experimentally accessible parameters of the system (shape, external forces, 
constraints, substrate features, contacts, etc.) so that any given desired PMF 
pattern can be generated under realistic experimental circumstances?

In this work we pose and address one such question. We demonstrate that the 
geometry of a substrate can be optimized so that a graphene Corbino-type device 
can have its conductance suppressed when hydrostatically pressurized. 
This, on the one hand, can lead to extremely high on/off ratios due to the two 
contrasting transport regimes in the pressurized and relaxed states. Pressurized 
devices are shown to leak current only through perfect, spatially separated 1D 
modes arising from snake-type states that are stabilized at the inversion 
regions of the PMF. This creates a very strong current pinch-off by squeezing 
(\emph{literally} in this case) the transport channel to the quantum wire limit 
and, in addition, establishes the possibility of having strictly quantized 
conductance on demand without electrostatic confinement. Moreover, since the 
chirality of PMF-induced snake states is uniquely tied to the valley degree of 
freedom, such devices can function as effective valley filters or sources of 
valley polarized currents.
Our calculations are performed in an optimization framework that tackles the 
``reverse'' strain-engineering problem in graphene. We begin with an overview 
of its main ingredients and a discussion of the optimal shapes for nearly 
uniform PMF throughout the Corbino ring. Subsequently, we describe how the 
deformation fields so obtained are mapped to an actual graphene lattice whose 
electronic properties are described in a non-uniform tight-binding 
approximation. This constitutes the basis for our study of the quantum transport 
characteristics and local electronic structure in devices of different 
dimensions, with and without disorder, whose results establish the behavior and 
phenomenology indicated above.

\section{Corbino shape optimization}
Stating the problem as an optimization question, we wish to find the shape of a 
non-circular annular cavity in a substrate such that the graphene flake, 
when deposited on it from above and subjected to hydrostatic pressure of 
magnitude $p$, exhibits a pseudomagnetic field (PMF) with a magnitude $B_0$ 
that is constant in as much area of the system as possible, compatible with 
realistic elastic constraints. The substrate shape and dimensions so determined 
would constitute the basic information for the design of the corresponding 
transport experiment.
If we assume that the graphene flake is rigidly attached to the substrate, 
finding the shape of the cavity is equivalent to finding the shape of the 
graphene flake with clamped boundary conditions at its edges, with the same 
optimized PMF under the applied hydrostatic pressure. This problem is solved by 
appealing to a Partial Differential Equation (PDE) constrained optimization
technique, discretized by finite 
elements \cite{JonesPereira2014}. It involves varying the control 
variables (the shape of the cavity) and the state variables (the plate 
deformation) so as to minimize a measure of the distance of the underlying PMF 
away from a certain desired PMF pattern, subject to the constraint that the 
state variables must satisfy the equations governing the elastic deformation of 
the graphene sheet for a given shape.

\subsection{Overview of the inverse optimization problem}
Our control variable is the shape of the domain occupied by the graphene 
flake (it is this that we numerically vary in order to approach the desired PMF 
in the pressurized field). 
We thus parametrize the domain using a set of control 
variables: $\Omega=\Omega[c_1,\ldots,c_n]$, to be specified later.
The clamped boundary conditions at the two perimeters (and beyond) where the 
graphene sheet contacts the substrate imply that the material domain 
of interest is the annular region defined by the patterning of the substrate, 
and not that part of the graphene that adheres to the substrate itself. 
The central mathematical problem is as follows:
\begin{align}
\textrm{Minimize}\quad{\mathcal{I}}= &  
\frac{1}{\area\Omega}\iint_{{\Omega}[c_1,\ldots,c_n]}({B}^2-B_0^2)^2\,\mathrm{d}
^2{\mb{X}} \nonumber \\
 + & \eta \, \mathcal{I}^{\textrm{reg}}[c_1,\ldots,c_n]
,\label{eq:opt}
\end{align}
subject to six constraint equations expressing the mechanical equilibrium of 
the graphene sheet, where the PMF $B$ is an explicit and known function of the 
strain field in the 2D sheet, and $B_0$ is the target PMF which, in general can 
be any predefined function of space. This formulation ensures that the $B$ 
sought in the optimization satisfies $B^2\approx B_0^2$ as closely as is 
feasible under the restrictions caused by the geometry and mechanics of the 
problem (the scheme seeks PMFs that are as close as possible to $\pm B_0$; in 
the \SI\ we discuss the case where the objective function is chosen to minimize 
$B-B_0$, rather than $B^2-B^2_0$).
$\Omega$ represents the domain occupied by the unadhered part of the graphene 
flake, defined by a characteristic lengthscale $L$. The explicit constraint 
equations that accompany \Eqref{eq:opt} are provided in the \SI{} in order not 
to obscure the main discussion 
here. The final term is a regularization term necessary for well-posedness of 
the optimization problem. In reality, the optimization problem will be solved 
in dimensionless form but we present the dimensional version here for clarity.

% ------------------------------------------------------------------------------
% FIGURE
% ------------------------------------------------------------------------------
\begin{figure*}[tb]
\subfigure[]{%
  \includegraphics[width=0.30\textwidth]{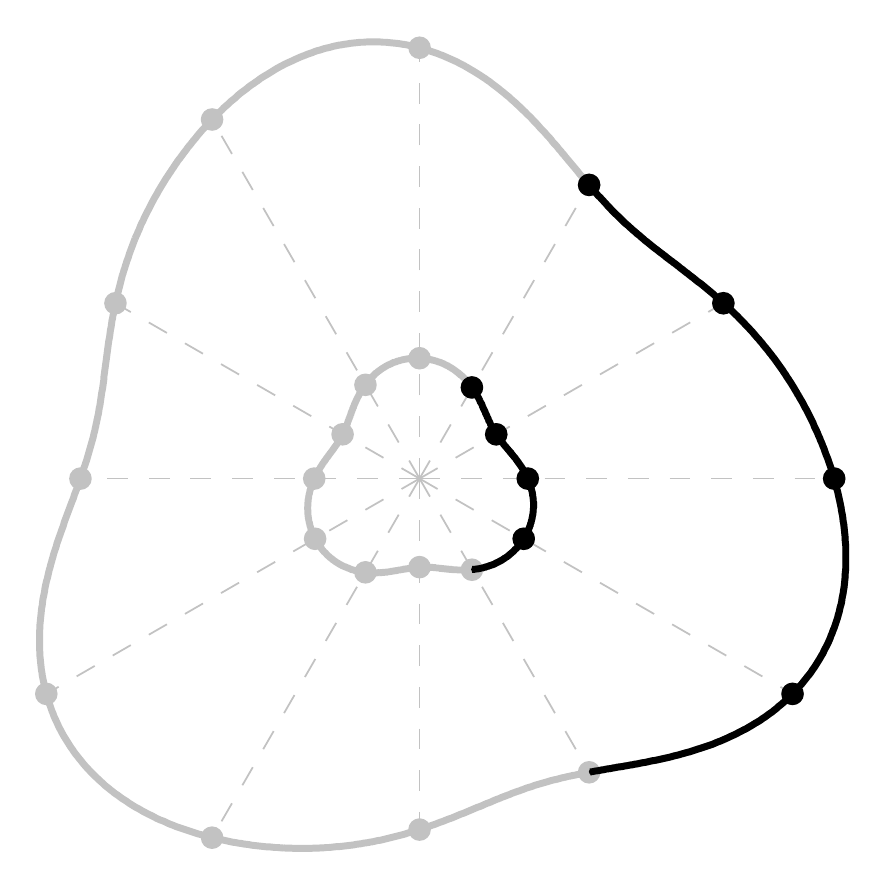}
  \label{fig:controlpoints}
  }
\subfigure[]{%
  \includegraphics[width=0.18\textwidth]{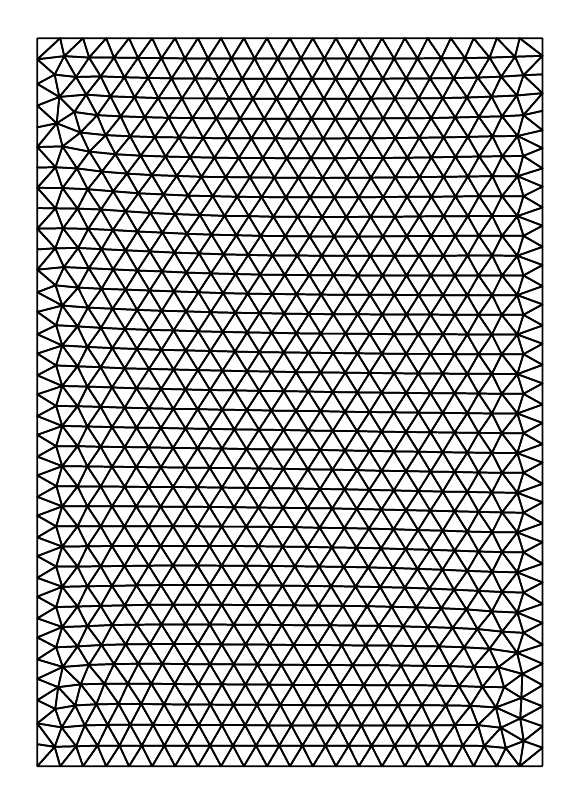}
  \label{fig:mesh1}
  }
\subfigure[]{%
  \includegraphics[width=0.14\textwidth]{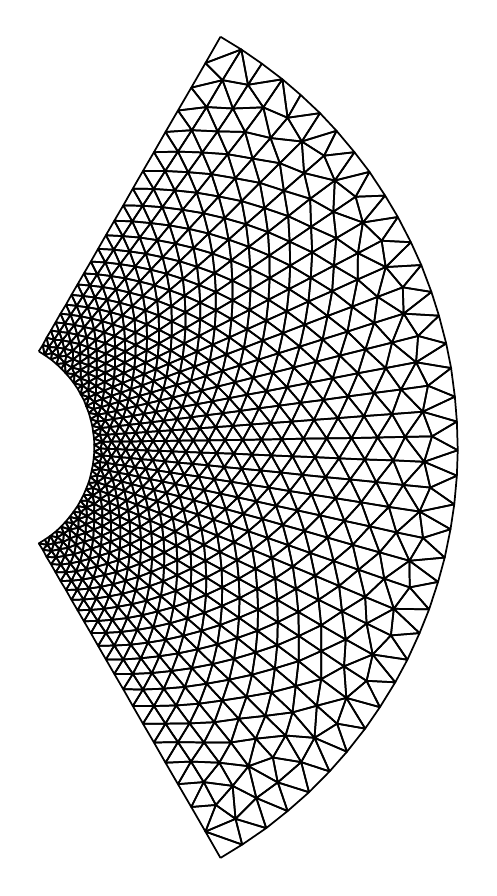}
  \label{fig:mesh2}
  }
\subfigure[]{%
  \includegraphics[width=0.30\textwidth]{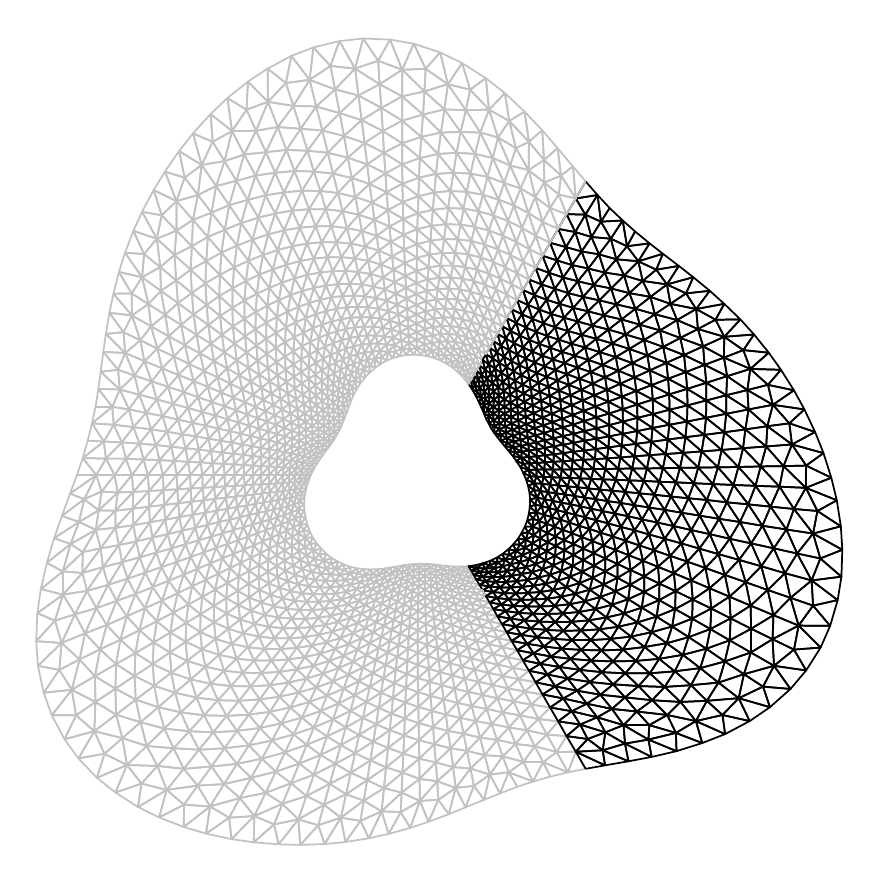}
  \label{fig:mesh3}
  }
\caption{
  \subref{fig:controlpoints}
  Construction of the domain shape from a discrete set of $N=4$ control points 
  for the inner and outer boundaries.
  \subref{fig:mesh1}
  A uniform grid on a rectangle. The $x$-coordinates of the upper and 
  lower boundaries must be in registration for the periodicity condition to 
  apply.
  \subref{fig:mesh2}
  The grid of \subref{fig:mesh1} has been mapped conformally to an 
  annular sector to increase the resolution at the inner boundary.
  \subref{fig:mesh3}
  The grid of \subref{fig:mesh2} has been stretched radially to match 
  the example shape shown in \subref{fig:controlpoints}. This mesh is repeated 
  twice to form a complete mesh, though only the sector in black is calculated.
  } 
\end{figure*}
% -----------------------------------------------------------------------------

\subsection{Discretization and numerical solution}
Each of the state variables (the three displacement components $v_1$, $v_2$, $w$ 
of the plate, and the three components of the bending moment $M_{11}$, 
$M_{12}$, $M_{22}$) is discretized using linear finite elements. This requires 
the generation of a triangular mesh covering the entire domain occupied by the 
graphene flake. However, the shape of this domain $\Omega$ must be parametrized 
by a finite set of variables in order for the optimization algorithm (which 
works in a discrete setting) to be appropriately formulated.
The simplest way to parametrize the shape of the domain $\Omega$ is to prescribe 
equidistant control points around the outer and inner boundaries, and 
interpolate them with a Fourier series in the angular coordinate $\theta$. 
Furthermore, due to the symmetry of the graphene lattice, the PMF is invariant 
if the coordinate axes are rotated by an angle of $2\pi/3$. Therefore we may 
restrict our domain to the region $-\pi/3\leq\theta\leq\pi/3$, and admit 
outlines of the form
\begin{equation}
R(\theta)=\sum_k(\alpha_k\cos3k\theta+\beta_k\sin3k\theta)
\end{equation}
only. An example of the domain construction from control points is shown in 
\Fref{fig:controlpoints}.
Let $N$ be the (even) number of control points on the outer boundary, with the 
same number on the inner boundary.
Then if $(R_i^{\textrm{out}})_{i=1,\ldots,N}$ are the radial coordinates of the 
outer boundary for the angular coordinates $\theta_i=2\pi(i-N/2)/(3N)$, we have
\begin{equation}
R_{\textrm{out}}(\theta)= \frac{\alpha_0}{2} 
+\sum_{k=1}^{N/2}\alpha_k\cos3k\theta 
+\sum_{k=1}^{N/2-1}\beta_k\sin3k\theta,\label{eq:radialFT}
\end{equation}
where the coefficients are found by solving 
$R_{\textrm{out}}(\theta_i)=R_i^{\textrm{out}}$ for $i=1,\ldots,N$.
The same procedure is repeated for the inner boundary control variables 
$R_i^{\textrm{in}}$.
The $2N$ control variables are allowed to vary within a user-provided interval,
\begin{equation}
L_{\textrm{min}}^{\textrm{in}}\leq R_i^{\textrm{in}}\leq 
L_{\textrm{max}}^{\textrm{in}},\qquad
L_{\textrm{min}}^{\textrm{out}}\leq R_i^{\textrm{out}}\leq 
L_{\textrm{max}}^{\textrm{out}}.
\end{equation}
For each set of control variables chosen during the iterative optimization 
process, we will have a domain of a different shape.
It would be improper to triangulate each new domain separately: not only would 
it be prohibitively time-consuming, but most optimization routines work best 
when the constraints and objective functions vary smoothly when the state and 
control variables are altered.
Therefore, we instead choose a mesh for an annular sector and deform the mesh 
as the control variables accordingly alter the domain. The annular sector is 
defined by $L_{\textrm{in}}\leq R \leq L_{\textrm{out}}$ and 
$-\pi/3\leq\theta\leq\pi/3$, where 
\begin{equation}
L_{\textrm{in}} 
=(L_{\textrm{min}}^{\textrm{in}} 
+L_{\textrm{max}}^{\textrm{in}})/2,\textrm{ and } L_{\textrm{out}} 
=(L_{\textrm{min}}^{\textrm{out}} 
+L_{\textrm{max}}^{\textrm{out}})/2.\label{eq:Rin_Rout}
\end{equation}
To mesh this annulus in a way which ensures a good resolution at the inner 
boundary, we construct a uniform mesh of $N_p$ points for the 
rectangle\footnote{A mesh size giving around 8 boundary elements per control 
point was used for the calculations in this 
article. Meshes were constructed using the Matlab-based mesh routine DistMesh 
\cite{PerssonStrang2004}.} 
\begin{equation}
\{(x,y):\log L_{\textrm{in}}\leq x\leq 0,\;\;-\pi/3\leq y\leq\pi/3\}.
\end{equation}
We conformally map the mesh for the rectangular domain to the annular sector 
using the mapping ${X}+\mathrm{i}{Y}=\exp(x+\mathrm{i}y)$.
Then, to deform this mesh to the domain defined by the control variables, we 
simply kept the $\theta$-component of the meshpoint fixed, and affinely 
displaced the radial component $R=\sqrt{{X}^2+{Y}^2}$ to
\begin{equation}
R_{\textrm{in}}(\theta)+(R_{\textrm{out}}(\theta)-R_{\textrm{in}}
(\theta))\left(\frac{R-L_{\textrm{in}}}{L_{\textrm{out}}-L_{\textrm{in}}}\right)
,
\end{equation}
where $R_{\textrm{out}}(\theta)$ and its equivalent at the inner boundary are 
calculated from the control points by (\ref{eq:radialFT}).
See \Fsref{fig:mesh1}--\ref{fig:mesh3} for a visualization of this 
procedure.
The purpose of the regularization term $\mathcal{I}^{\textrm{reg}}$ in the 
objective function (\ref{eq:opt}) is to make the optimization problem well-posed 
by penalizing solutions that have unfavorable properties. In this study  we set 
the regularization function to be 
\begin{equation}
\mathcal{I}^{\textrm{reg}}=\int_{-\pi/3}^{\pi/3}(R_{\textrm{out}}''(\theta)^2+R_
{\textrm{in}}''(\theta)^2)\,\mathrm{d}\theta.
\end{equation}
in order to penalize highly-convoluted boundaries.

As discussed previously, the state variables are ${v}_1$, ${v}_2$, ${w}$, 
${M}_{11}$, ${M}_{12}$, ${M}_{22}$, and are discretized such that the equations 
are written in terms of their values at each node in the triangulation. 
Additionally there are boundary conditions to be applied:\ the displacements are 
set to zero on nodes comprising the inner and outer boundaries. At the 
boundaries $\theta=\pm\pi/3$, we impose rotated periodicity conditions on these 
quantities (expressed in polar coordinates $R$, $\theta$), \emph{i.e.}\ 
$w|_{\theta=\pi/3}=w|_{\theta=-\pi/3}$, and similarly for $v_R$, $v_\theta$, 
$M_{RR}$, $M_{R\theta}$, $M_{\theta\theta}$.
The discretized PDE system becomes a discrete optimization system:\ to minimize 
an objective function $F(\mb{C},\mb{U})$ subject to $6N_p$ constraints 
$\mb{G}(\mb{C},\mb{U})=\mb{0}$, where $\mb{C}$ is the vector of control 
variables and $\mb{U}$ is the vector of $6N_p$ state variables.

\section{Optimized device shapes}
\begin{figure*}[tb]
\subfigure[]{%
  \includegraphics[height=0.3\textwidth]{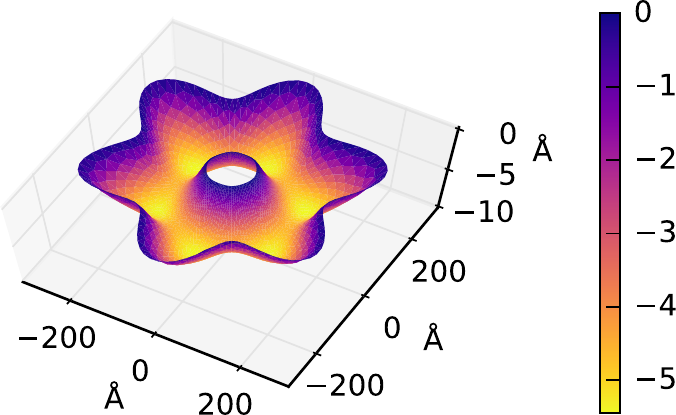}
  \label{fig:flower-inflated}
  }
  \qquad
\subfigure[]{%
  \includegraphics[height=0.3\textwidth]{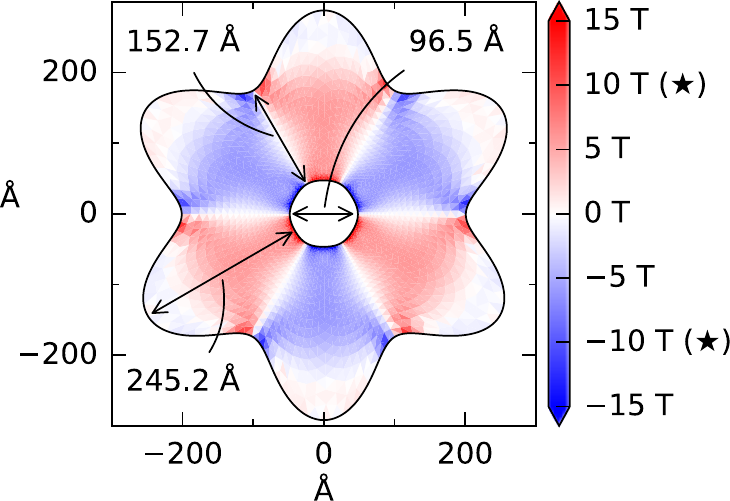}
  \label{fig:flower-pmf}
  }
\caption{(color online)
\subref{fig:flower-inflated} An elevation plot (with exaggerated 
vertical scale) of the pressurized `flower'-shaped solution. Colors indicate 
elevation, to a largest value $5.5\unit{\AA}$. 
\subref{fig:flower-pmf} 
The 
corresponding PMF. Target PMF was $B_0=10\unit{T}$ (indicated by $\bigstar$ on 
the colorbar). Characteristic lengthscales are indicated on the diagram.
}
\label{fig:flower}
\end{figure*}

As the electronic and mechanical properties of graphene remain fixed, there 
are only three parameters that we can change in this system: the limits on the 
annulus shape (and hence the lengthscale $L$), the applied (constant) pressure 
$p$, and the target PMF $B_0$. In our calculations we allowed the $N=4$ outer 
control points to vary in a range $261\unit{\AA}\pm 60\unit{\AA}$ and the inner 
points in a range $61\unit{\AA}\pm 15\unit{\AA}$. We initially set a target PMF 
of $B_0=10\unit{T}$ and the pressure was fixed at 
$100\unit{bar}=10^7\unit{Pa}$. Writing the coefficients to the nearest 
$0.1\unit{\AA}$, the shape of the optimal outer and inner boundaries were 
calculated to be (in Angstroms)
\begin{align}
R_{\textrm{out}}(\theta) & =
  245.4+0.0\cos3\theta-44.4\cos6\theta+2.0\sin3\theta, 
\label{eq:flower_out}\\ 
R_{\textrm{in}}(\theta) & = 
  47.6+0.1\cos3\theta+0.6\cos6\theta-0.3\sin3\theta. 
\label{eq:flower_in}
\end{align}
This outline corresponds to a striking flower-like geometry, and its deformation 
under hydrostatic pressure is displayed in \Fref{fig:flower-inflated}. To this 
deformation corresponds the spatial distribution of PMF shown in 
\Fref{fig:flower-pmf}. Though the target PMF was $10\unit{T}$, the 
root-mean-squared value of the calculated PMF was only $4.05\unit{T}$. It should 
be kept in mind that the target PMF is not in general attained in these 
optimization calculations, but is rather a mechanism for forcing the solution 
towards our desired direction.

% ------------------------------------------------------------------------------
% FIGURE
% ------------------------------------------------------------------------------
% ------------------------------------------------------------------------------

\begin{figure*}[tb]
\includegraphics[width=1\textwidth]{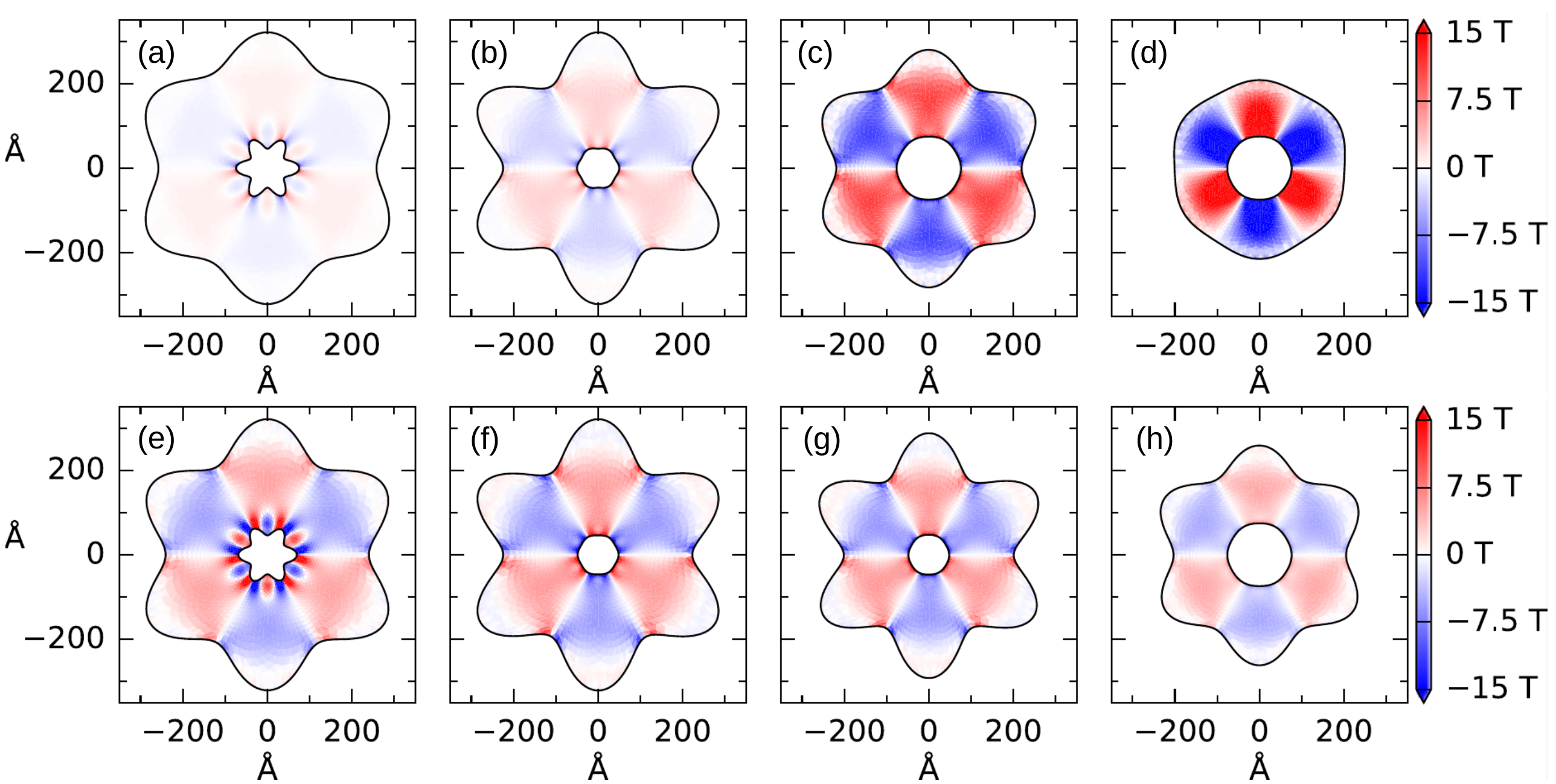}
\caption{(color online)
Upper row (a--d): comparison of the shape and PMF distribution of devices 
optimized for a target $B_0{\,=\,}10\unit{T}$ and fixed 
$\kappa{\,=\,}5.76\times10^{-5}$. Each panel is the result of the optimization 
at a different pressure: $p{\,\in\,}\{10\unit{bar}, \,10\sqrt{10}\unit{bar}, 
\,100\sqrt{10}\unit{bar}, \,1\unit{kbar}\}$ (this corresponds to the set of 
dimensionless pressures $\bar{p}{\,\in\,}\{0.0337,\,0.107,\,1.07,\,3.37\}$).
Lower row (e--h): result of the optimization for different target PMFs of 
$50\,\textrm{T}$, $20\,\textrm{T}$, $10\,\textrm{T}$, and $4\,\textrm{T}$. 
Pressure was kept fixed at $100\unit{bar}$ and $L=261\unit{\AA}$. 
In all panels (a--h) we used $L{\,=\,}261\unit{\AA}$.
}
\label{fig:pbar_variation_flowerB0}
\end{figure*}

Since solutions to the optimization problem in \Eqref{eq:opt} are obtained for 
a specific choice of input parameters $p$ and $B_0$, one immediate and 
important question is to assess how robust is the solution for the shape with 
respect to changes in those parameters.
As the optimization and its constraints were solved in a dimensionless setting, 
the behavior of the solution is dependent on the values of dimensionless 
parameters rather than true physical quantities. The nondimensionalized system 
has only two parameters: the dimensionless bending stiffness $\kappa$, and the 
dimensionless pressure $\bar{p}$. In terms of the three physical quantities $p$, 
$L$, and $B_0$, these parameters behave as (details in the \SI)
\begin{equation}
\bar{p}\propto\frac{p}{B_0^{3/2}L^{1/2}},\qquad \kappa\propto\frac{1}{L^3B_0}.
\label{eq:dimlesspar}
\end{equation}
Varying $\kappa$ has little effect on the shape obtained through the 
optimization procedure (see Fig.\ S1 and related discussion in the \SI), and so 
only changes in $\bar{p}$ are relevant in understanding how the optimized shape 
depends on the choices of input parameters. For instance, increasing 
the hydrostatic pressure $p$ by a factor of $8$ and the target PMF $B_0$ by a 
factor of $4$ has no effect on the optimized shape, as the dimensionless 
constant $\bar{p}$ is kept unchanged and the effect of a changed $\kappa$ is 
negligible. In contrast, if $\kappa$ is kept constant and $\bar{p}$ is allowed 
to vary, we see a significant variation in the calculated shape as shown in 
\Fsref{fig:pbar_variation_flowerB0}a--d.
This variation in shape with changes in $\bar p$ deserves further explanation. A 
change in $\bar p$ corresponds to varying $p$ while keeping the lengthscale and 
target PMF constant. But a better way of seeing the effect is to think about an 
increase in $\bar p$ as a decrease in target PMF $B_0$, while keeping $p$ and 
$L$ constant [cf. \Eqref{eq:dimlesspar}]. Though this also changes the value of 
$\kappa$, we have noted that the effect of this is minor.
To explore this line of thinking, in the lower row of 
\Fref{fig:pbar_variation_flowerB0}, we report a further set of calculations 
which kept $p=100\unit{bar}$ and $L=261\unit{\AA}$ fixed while varying the 
target PMF $B_0$ through $50\,\mathrm{T}$, $20\,\mathrm{T}$, $10\,\mathrm{T}$, 
and $4\,\mathrm{T}$. For high target PMFs, the optimization forces the inner 
boundary to be highly convoluted in order to obtain the high strain gradients 
needed to achieve the target PMF. For low target PMFs, an annular solution is 
already close to the optimal solution, and the optimization only needs to 
make small adjustments to the outline to make the magnitude of the field more 
uniform. The results for $10\unit{T}$ and $20\unit{T}$ can be seen as 
intermediates between these two extremes. The two extremes also explain the 
results of \Fref{fig:pbar_variation_flowerB0}a--d calculated using different 
pressures for a given $B_0$:\ for a low pressure, a convoluted interior 
boundary is required to hit the target PMF, while for high pressure an annular 
solution is already close to the target and only minor modifications of shape 
are needed.
In the end, the two rows of \Fref{fig:pbar_variation_flowerB0} illustrate what 
is implied by \Eqref{eq:dimlesspar}: the changes in the optimal shape with 
increasing $p$ (and fixed $B_0$) follow the same trend as the changes with 
decreasing $B_0$ (and fixed $p$).
Of the calculations shown in \Fref{fig:pbar_variation_flowerB0}e--h we 
selected the solution (g), which was shown before in \Fref{fig:flower}, 
with $B_0=10\unit{T}$ and $\bar p = 0.337$ as giving a good middle ground 
between PMF smoothness and strength.

% ------------------------------------------------------------------------------
% FIGURE
% ------------------------------------------------------------------------------
% ------------------------------------------------------------------------------

If we are now given an arbitrary lengthscale $L$ and pressure $p$, in order to 
find the optimal shape we need to choose an appropriate value of $B_0$.
Since we have seen that the result for $\bar{p}=0.337$ gives an acceptable 
outcome, we could choose a $B_0$ that results in this value of $\bar{p}$. This 
gives us a system of equations with the same value of $\bar{p}$ but a different 
value of $\kappa$ as our first calculation. We have noted above that varying 
$\kappa$ has little effect on the shape of the outline at constant $\bar{p}$. 
So, as long as the new value of $\kappa$ does not fall too much beyond the range 
of magnitudes where we verified this to be true (supplementary Fig.\ S1), the 
solution will be similar to that in equations (\ref{eq:flower_out}, 
\ref{eq:flower_in}), plotted in \Fref{fig:flower-pmf}.
Since the differences are small, and would be dwarfed by manufacturing 
variability in a real system, we kept the same shape 
(\ref{eq:flower_out})--(\ref{eq:flower_in}) in all our atomistic transport 
calculations to be discussed below for the sake of comparability. This shape is 
henceforth referred to as the \emph{flower device}. The initial optimization was 
for a pressure of $p=100\,\textrm{bar}$, and we performed further 
pressurizations of the shape up to $5\,\textrm{kbar}$, for which the 
distribution of PMF was not greatly changed, though its magnitude was quite 
different. Each of these pressurizations is identified below by the maximum 
strain achieved in the graphene sheet, up to a maximum strain of 
$\varepsilon_m = 6.11\%$ (note, however, that, even though our analysis is based 
on a set of specific parameter values, the scaling relations 
\eqref{eq:dimlesspar} allow the freedom to explore a large range of relevant 
pressures, target PMFs, or system sizes without having to run a new shape 
optimization procedure for each particular choice; see, for example, Fig.\ S6 in 
the \SI).

\section{Transport characteristics}\label{sec:transport}
To access the transport properties of the flower device we use the 
Landauer--B\"uttiker formalism; specifically, we are interested in the quantum 
conductance of the device, which is calculated by Caroli's formula \cite{Meir} 
$G=\frac{2e^2}{h} \Tr[\Gamma_qG^r\Gamma_pG^a]$, where 
$G^r=[G^a]^{\dagger}=[E+i\eta-H-\Sigma_p-\Sigma_q]^{-1}$ is 
the retarded [advanced] Green's function, the coupling between the contacts and 
the device is represented by $\Gamma_q=i[\Sigma_q-\Sigma_q^{\dagger}]$, and 
$\Sigma_q$ is the self-energy of the contact $q$. The geometry of the device is 
easily included in a nearest-neighbors $\pi$-band tight-binding Hamiltonian by 
merely selecting the sites that are located between the mathematically defined 
inner and outer edges [cf.\ \Fref{fig:flower-pmf}]. 
The inner contact is essentially a circular contact with $R_{\textrm{in}} 
\approx 47\unit{\AA}$; the outer contact radius varies between 
$200\unit{\AA}\leq R_{\textrm{out}}\leq 322\,\unit{\AA}$ with the angular 
modulation prescribed by \Eqref{eq:flower_out}. The calculation of 
the conductance requires the Green's functions of these non traditional 
contacts; this problem can be resolved by recalling that, for graphene, the 
particular details of the contacts can be replaced by an effective self-energy 
term provided that the contacts inject a high number of modes (highly doped 
contacts) \cite{yo_corbino}. Under this model, an effective self-energy term 
$-i|t|$ is added to the onsite energy of the atoms at the edges ($t = 
-2.7\unit{eV}$ is the graphene hopping parameter).  

The mechanical deformations induced by the hydrostatic pressure are 
incorporated in the tight binding Hamiltonian through the modification of the 
hopping parameter between neighboring sites as \cite{kirigami} 
\begin{align}
t_{ij}= & \, V_{pp\pi}(d_{ij}) \, \widehat{n}_i \cdot \widehat{n}_j \nonumber 
\\ 
& + \bigl[ V_{pp\sigma}(d_{ij})-V_{pp\pi}(d_{ij}) \bigr]
  \frac{(\widehat{n}_i \cdot \vec{d}_{ij}) (\widehat{n}_j \cdot \vec{d}_{ij}) 
}{d^2_{ij}},
\label{eq:var_t}
\end{align}
where $\widehat{n}_i$ is the vector normal to the surface at point $i$,
$\vec{d}_{ij}$ is a vector linking the two sites, and $V_{pp\sigma}$ and
$V_{pp\pi}$ are the Slater--Koster parameters, modified to account for changes 
in the bond length through $V_{pp\pi}(d_{ij})=t\, e^{-3.37(d_{ij}/a-1)}$ and
$V_{pp\sigma}(d_{ij})=-2.4V_{pp\pi}(d_{ij})$  \cite{Val_SK,Vmudt,Vpps}.

% ------------------------------------------------------------------------------
% FIGURE
% ------------------------------------------------------------------------------
\begin{figure*}[tb]
\centering
\includegraphics[width=\textwidth]{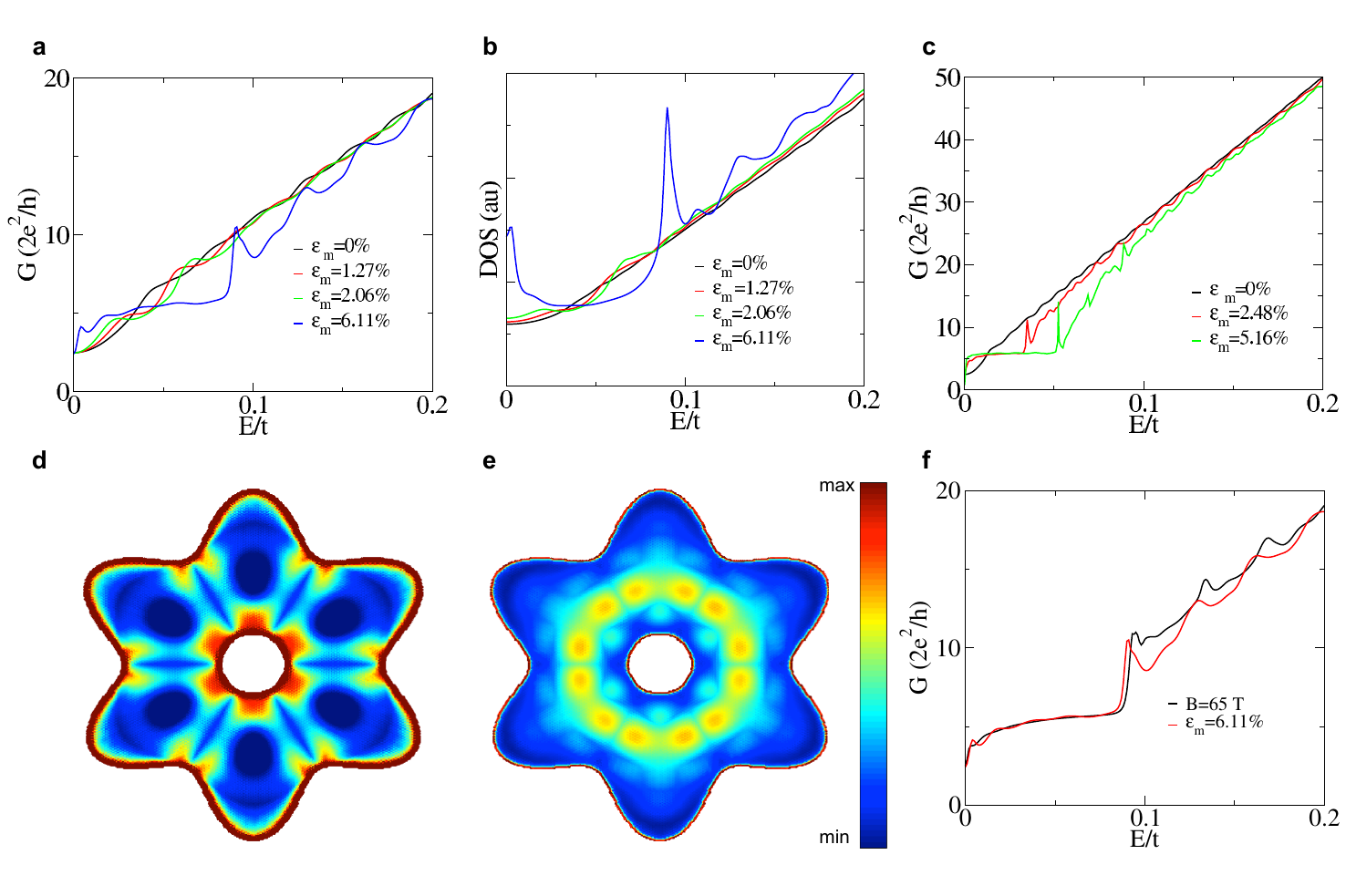}
\caption{(color online)
The top row shows plots of the conductance (a) and the DOS (b) of a device with 
$R_{\textrm{in}}\approx4.7\unit{nm}$ and $20\unit{nm}\leq 
R_{\textrm{out}}\leq32.2\unit{nm}$ ($\alpha=1$) 
for different values of maximal strain $\epm$, as well as the conductance of a 
similar device scaled with $\alpha=3$ for different values of maximal strain 
(c).
The bottom row shows the LDOS at $E=0.04t$ (d) and $E=0.09t$ (e) of the device 
in (a). (f) Comparison of the conductance of the flower device used in (a) 
for $\epm=6.11\%$ with that of an unpressurized one, where the latter includes 
a real magnetic field ($B=65\unit{T}$) with a simulated spatial pattern 
that matches the one of the PMF [see also \Fref{fig:fig2_te}a].
}
\label{fig:te_flor_str}
\end{figure*}
% ------------------------------------------------------------------------------

To best appreciate the features that arise from the pressure-induced PMF, it is 
useful, first, to have a perspective over the main transport characteristics of 
unpressurized devices which define our scenario of reference; the results of 
these calculations are provided in the \SI. 
Devices were generated from the optimized flower shape shown in 
\Fref{fig:flower-pmf} ($R_{\textrm{in}}\approx47\unit{\AA}$ and 
$200\unit{\AA}\leq 
R_{\textrm{out}}\leq322\unit{\AA}$). \Fref{fig:te_flor_str}a shows the 
conductance of pressurized flower devices for different values of the maximal 
strain in the graphene flake, $\epm$, that we use to distinguish different 
cases. Notably, at $\epm\approx 6\,\%$, the conductance starts developing a 
plateau of $5.7(2e^2/h)$ with a resonant peak at $E=0.09t$; the plateau occupies 
the whole energy range $0\lesssim E \lesssim 0.09t$. When the corresponding 
density of states (DOS) is analyzed [cf. \Fref{fig:te_flor_str}b], one 
identifies two  sharp peaks at $E=0$ and $E = 0.09\,t$ arising from the 
formation of strain-induced LL with $n=0,\,1$. From these, we extract a PMF 
$B_s\approx 65\unit{T}$ and this estimate allows us to further confirm that 
local maxima occurring in the conductance of \Fref{fig:te_flor_str}a correlate 
to higher energy LL ($E_2=0.13t$, $E_3=0.16t$ and $E_4=0.18t$). 
The dependence of the conductance on device size for the same spatial 
distribution of PMF can be studied without having to run a new shape 
optimization procedure by exploring the scaling implied by 
\Eqref{eq:dimlesspar}. For example, if one scales the inner and outer 
boundaries as $R^{\alpha}_{\textrm{in}}(\theta)=\alpha 
R_{\textrm{in}}(\theta)$, $R^{\alpha}_{\textrm{out}}(\theta)=\alpha 
R_{\textrm{out}}(\theta)$, these relations can be used to extract the pressure 
needed to achieve the same PMF pattern, as well as the magnitude of the 
quasi-uniform field (see Fig.~S5 and related discussion in the \SI).
The conductance of a device thus scaled with $\alpha=3$ ($R_{\textrm{out}}\simeq 
80\unit{nm}$) is shown in \Fref{fig:te_flor_str}c for maximal strains of 
$\epm=2.48\%$ and $\epm=5.16\%$. In the two cases both the plateau and resonant 
peak are again present, from which we estimate $B_s \approx 9\unit{T}$ and $B_s 
\approx 20\unit{T}$, respectively. However, in larger devices these features are 
much more sharply defined and the plateau is flatter: the conductance is 
therefore more perfectly quantized at $6(2e^2/\hbar)$ with growing device 
dimension. They also require less strain to emerge due to the fact that the 
inner and outer contacts become decoupled at lower PMF (because the condition 
that the magnetic length, $\ell_B{\,=\,}\sqrt{\hbar/e B}$, is much less than $L$ 
is met at smaller PMF). 

\subsection{Perfectly conducting channels and valley filtering}
In order to understand the origin of this robust quantization induced by 
pressure over a large range of energies that are experimentally relevant, let us
look first at the real-space distribution of the electronic wavefunctions. The 
local DOS (LDOS) calculated at the plateau midpoint ($E=0.04t$) and at the peak 
($E=0.09t$) are shown in \Fsref{fig:te_flor_str}d and 
\ref{fig:te_flor_str}e, respectively, for the device with $\alpha=1$ 
pressurized to $\epm=6.11\%$ [cf. \Fref{fig:te_flor_str}a].
At the peak [$E=0.09t$, \Fref{fig:te_flor_str}e], the observation of a state 
entirely confined in the central portions of the device identifies it as 
clearly associated with one of the PMF-induced LLs \cite{Qi:2013}. 
In sharp contrast, for energies in the plateau [$E=0.04t$, 
\Fref{fig:te_flor_str}d] the LDOS concentrates on six narrow radial channels 
of the flower device. Comparing the LDOS with the PMF map shown in 
\Fref{fig:flower-pmf} reveals that the wavefunction concentrates at 
precisely\,---\,and only\,---\,those regions where the PMF changes polarity. 
Consequently, the plateau in conductance at low energies is associated with 
current being carried by these pseudomagnetic interfacial states, similarly to 
the corresponding situation in nonuniform (real) magnetic fields 
\cite{Park:2008yu} where electrons propagate chirally along a boundary 
separating fields of opposite polarity.

A useful picture of the nature of these 1D modes stabilized at the polarity 
inversion interfaces is provided by the semi-classical limit of this problem 
where they become so-called snake states \cite{snake}. The designation 
arises from the different winding sense of the electron orbits in regions where 
$B$ has opposite polarity, which allows them to be confined and propagate along 
the interface with a definite direction. This interpretation helps one to 
understand: (i) directionality of the states and (ii) localization in regions 
where $B_s \approx 0$ and changes sign. However, despite the usefulness of 
the semi-classical perspective, it is important to retain that no clear 
correspondence to the classical motion can be established in the case shown in 
\Fref{fig:te_flor_str}d because, as we are looking at energies between 
the $n=0$ and $n=1$ LL, the cyclotron radius is smaller than the magnetic length 
for the energies studied: $r_c = \ell_B^2 k_F \sim \ell_B / \sqrt{2} < \ell_B$ 
\cite{Rickhaus:2015kx,Taychatanapat:2015nr}; these modes are in the quantum 
regime.  

% Note: the second step in $r_c = \ell_B^2 k_F \sim \ell_B / \sqrt{2}$ is done
% by putting \hbar k_F = E_1 / 2 (the midpoint between the n=0 and n=1 LLs).

\begin{figure*}[tb]
\centering
\includegraphics[width=0.65\textwidth]{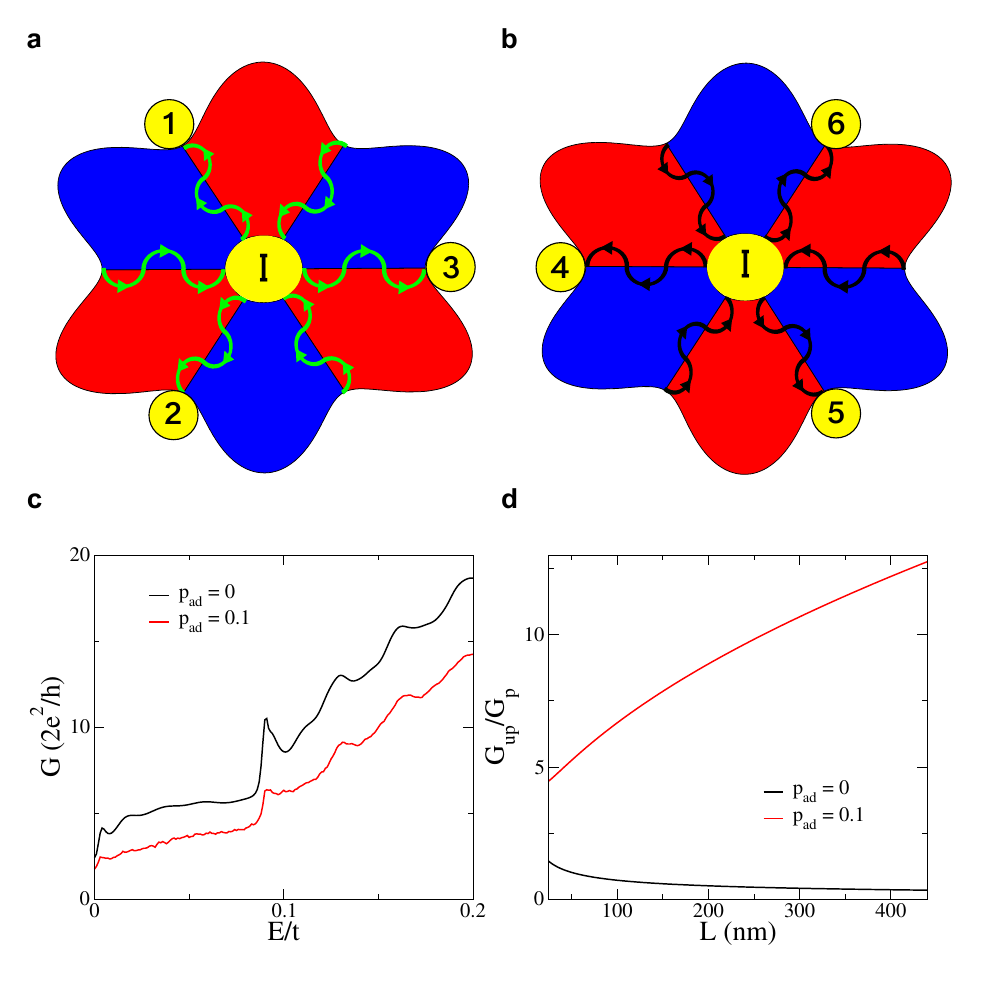}
\caption{(color online)
Schematic distribution and polarity of $B_s$, as well as the direction 
of propagation of the snake states, for electrons in valley $K$ (a) valley $K'$ 
(b). In our device, 
transport occurs via transmission from the inner contact (I) to the six outer 
contacts labeled $1\dots6$, where 1--3 collect electrons from valley $K$ only 
(panel a), and 4--6 from valley $K'$ (panel b).
(c) Conductance of the pressurized $\alpha=1$ flower device when the 
probability of adsorbing adatoms onto graphene is $p_{ad}=0.0$ (clean) and 
$p_{ad}=0.1$ 
(disordered).
(d) The unpressurized/pressurized conductance ratio ($\Gup/\Gp$) as a function 
of the device's linear dimension ($L$) for $p_{ad}=0.0$. This ratio is computed 
at the midpoint between the 0th and 1st Landau level energies of the 
pressurized device.
}
 \label{fig:fig2_te}
\end{figure*}

At low-energies, transport takes place via one-dimensional perfectly conducting 
modes, which is entirely consistent with the 1D-like energy dependence of the 
DOS plotted in \Fref{fig:te_flor_str}b for $\epm=6.11\%$. Moreover, the six 
neighboring sectors of alternating PMF polarity in our flower device are 
expected to beget six snake state channels, so one may deduce the existence of a 
plateau in the conductance of $6(2e^2/h)$, which is very approximately the case 
seen in \Fsref{fig:te_flor_str}a and \ref{fig:te_flor_str}c. 
Note that, while one might think that the difference between the observed 
[$5.7\,(2e^2/h)$] and theoretical values [$6\,(2e^2/h)$] of the conductance 
plateau is due to the inhomogeneity of PMF over each sector, a direct 
comparison between the  conductance of the pressurized device with an 
unpressurized flower under a real magnetic field with artificially sharp 
polarity changes shows no difference.  
More specifically, we have calculated the conductance of the same unpressurized 
device with an external magnetic field of constant magnitude ($B=65\unit{T}$) 
but alternating in sign at the six places where the PMF does so [cf. 
\Fref{fig:flower-pmf} and \Fref{fig:fig2_te}a] (additional details and 
different cases are discussed in the \SI). \Fref{fig:te_flor_str}f 
demonstrates that the conductance of the pressurized device with $\epm =6.11\%$ 
is essentially the same as that of the idealized unpressurized device with real 
magnetic field of the same magnitude. The agreement is nearly perfect in the 
plateau region below the first LL, further corroborating that the flower device 
is the optimal solution (to emphasize this point, in the supplementary Fig.\ 
S4(e) we show that the conductance of pressurized annular devices that retain a 
perfectly circular shape does not develop the quantization plateau). 
It also validates our earlier argument regarding the origin of the chiral 1D 
channels in the pressurized devices and their close analogy to the 
semi-classical snake states. 
In the end, we can trace the small deviation from the ideal $6\,(2e^2/h)$ 
quantization in the pressurized devices of \Fref{fig:te_flor_str}a to a small 
coupling between the modes belonging to different radial segments: to avoid 
such coupling requires $R_{\textrm{in}} \gtrsim 2\ell_B$ so that the wave 
functions of neighboring modes do not overlap near the inner contact, whereas 
this device has $R_{\textrm{in}}\approx 4.7\unit{nm}$ and $\ell_B = 
3.1\unit{nm}$ which falls short of it. In the supplementary Fig.\ S4 we show 
that perfect quantization is indeed achieved when $R_{\textrm{in}}$ and 
$\ell_B$ fulfill that criterion and that, as expected more generally, the 
conductance step is quantized at $G=2n_s\,e^2/h$, where $n_s$ is the number of 
polarity changing interfaces in the disk.

Given this scenario, the way in which PMFs couple to the electronic motion in 
graphene has an interesting and useful implication. Since electrons belonging 
to distinct valleys feel PMFs with opposite sign 
\cite{Suzuura2002Phonons,Vozmediano2010Gauge}, they propagate in opposite 
directions along the polarity boundary of the field. This effect is  
schematically illustrated in \Fsref{fig:fig2_te}a for an electron from valley 
 $K$, and \ref{fig:fig2_te}b for valley $K'$. 
Looking back at \Fref{fig:te_flor_str}d, each radial segment along which the 
LDOS is peaked supports two chiral 1D modes, each propagating in opposite 
radial directions, and each associated with one given valley.
Suppose then that the outer contact does not cover the entire perimeter but is, 
instead, sectioned into six outer contacts that allow collection of 
current only from the vicinity of the 1D channels (for our geometries that 
would mean defining contacts in the vicinity of the indentations of the 
outer perimeter, as in \Fsref{fig:fig2_te}a and \ref{fig:fig2_te}b). Under a 
given bias between inner and outer contacts (say, electrons flowing radially 
outward), the electrons arising from valley $K$ would reach three specific 
contacts (all equivalent, oriented $2\pi/3$ apart from each other), whereas 
those associated with valley $K'$ would reach the other three contacts, as 
indicated in \Fsref{fig:fig2_te}a and \ref{fig:fig2_te}b. In this schematic, 
electrons from valley $K$ are channeled by the 1D modes from the inner contact 
``I'' to the outer contacts 1, 2 and 3, while those from valley $K'$ can only 
reach contacts 4, 5 and 6.
A change in the bias sign, or changing from electron to hole-doped graphene with 
a back gate, would exchange the valleys that ``reach'' a given contact. In this 
manner, the device can spatially separate individual contributions from each 
valley and deliver valley polarized current to specific contacts or, 
alternatively, selectively filter or probe the existence of valley polarized 
currents.

We highlight that the observation of these effects in our flower-like geometry 
is not limited to the specific range of hydrostatic pressure and device 
dimensions used for the calculations reported in \Fref{fig:fig2_te}. 
For example, using the scaling relations (\ref{eq:dimlesspar}), we expect that 
scaling up the device to $L=400 \unit{nm}$ while preserving its shape and using 
a pressure of $\sim 5.5 \unit{bar}$ shall create a PMF of $\sim 0.26 \unit{T}$ 
with the same spatial distribution. The valley filtering effect will take place 
as long as LLs can be formed in the center of each petal guaranteeing that 
transport only takes place via the 1D modes. A simple estimate of the electronic 
mobilities needed for this can be made from the criterion that the mean free 
path, $\ell$, be larger than the characteristic cyclotron diameter: $\ell 
\gtrsim 2 \ell_B$. In terms of the mobility defined as $\mu{\,=\,}\ell e/\hbar 
k_F$ and using $k_F{\,=\,}\sqrt{2}/l_B$ for electrons in the first LL, this 
becomes $\mu \gtrsim 2 \ell_B e / \hbar k_F {\,=\,} \sqrt{2} \ell_B^2 e / \hbar 
{\,\sim\,} 5{\times}10^4\,\text{cm}^2/\text{V\,s}$, where we used $l_B{\,=\,} 
\sqrt{\hbar/eB} {\,\approx\,} 50 \unit{nm}$; such mobilities are rather 
standard in graphene devices (see Fig.~S6 and related discussion in the \SI\ for 
additional details).

Recent research has shown the existence of snake states in strained  graphene 
nanoribbons \cite{,PhysRevLett.112.096805,PhysRevB.94.075432}, and diverse 
valley filtering devices have been proposed by exploiting strained graphene 
\cite{PhysRevB.94.125422,doi:10.1021/nl1018063,PhysRevLett.106.136806,1.3473725, 
1608.04569,1.4967977}. Similarly to ours, these devices require a specific 
geometry. However, ours does not require external electric or magnetic fields,
produces valley filtered currents equally for both valleys, and whether the 
filtering occurs or not can be controlled by pressure or strain. 

% ------------------------------------------------------------------------------
% FIGURE
% ------------------------------------------------------------------------------
% ------------------------------------------------------------------------------

\subsection{Effects of disorder and high on/off ratios}
As often happens, disorder can simultaneously be detrimental or functional in 
solid state systems. From the first point of view, the conductance plateau 
associated with the snake states should be affected by disorder in real 
samples. As each radial channel supports one pair of opposite chirality and 
opposite valley quantum number, short-range scatterers will be particularly 
disadvantageous since they cause inter-valley scattering and, because of the 
valley-chirality locking, that translates into backscattering and the 
degradation of their perfectly conducting nature.
Additionally, the snake states' locations coincide with the sites of highest 
curvature in the pressurized flower where the probability of adsorbing alien 
atoms or molecules is higher \cite{Geim_status,Cretu2010Migration}. In order to 
assess this in a specific manner, we modeled the attachment of adatoms to 
graphene in these regions by adding an Anderson-impurity type perturbation to 
the electronic Hamiltonian: each carbon can bind an adsorbate with probability 
$p_{ad}$; this introduces a new electronic level of energy $E_{ad}=t/16$ 
that the electron can hop to from the attached carbon with a hopping amplitude 
$V=2t$ that describes the degree of hybridization \cite{Peres}.
The conductance averaged over an ensemble of 20 such systems with $p_{ad}=0.1$ 
for the device of \Fref{fig:te_flor_str}a is plotted in \Fref{fig:fig2_te}c. 
As anticipated, the presence of these dopants in the regions where the snake 
states occur increases backscattering, thus lowering the conductance. The 
plateau is seen to remain, but its height depends inversely on the size of the 
system (not shown); in this particular case, traces of quantization are still 
present due to the small width of the device used ($\Weff \approx 223 
\unit{\AA}$). 
We point out that the approximate preservation of the plateau 
under strong disorder (it affects 10\,\% of the atoms) is further evidence of 
the 1D character and resilience of the underlying modes: there is now 
backscattering (the plateau has smaller $G$), but the one-dimensionality is 
unaffected (a plateau largely remains).

Whereas such a system with strong short-range scattering will be less effective 
as a valley filter, the sensitivity of the low-energy base conductance to the 
radial dimension can have practical functional applications, as we now describe.
Under pressure, the perfectly conducting channels limit the minimum conductance 
that can be achieved, irrespective of how large the magnitude of the PMF might 
be. From this standpoint, they are leaky electronic devices without an off 
state. But the residual conductance can be controlled by the type and amount of 
disorder, suggesting a tuneable on-off ratio.
To simulate larger devices without incurring a high computational cost, and 
given that all of the above establishes the one-dimensional character of 
transport at low energies, we resort to a simple one-dimensional model 
\cite{Datta} to scale up the conductance for larger pressurized disordered 
devices: 
\begin{equation}
  \Gp(\Weff)=6\mathcal{T}\,\Bigl(\frac{2e^2}{h}\Bigr)
  , \qquad \mathcal{T}=\frac{T}{T+(1-T)N}
  .
\end{equation}
This assumes that in the snake-type 1D modes, electrons transmit through one 
adatom with probability $T$ and $N=\left(\Weff/a\right)p_{ad}$  is the number 
of scatterers in the channel of width $\Weff$. It follows that wide devices 
should have a notorious reduction of the conductance, entailing high ratios of 
unpressurized/pressurized conductance ($\Gup/\Gp$) in the energy region of the 
snake states. This ratio can be used as a figure of merit directly related to 
the on-off ratio of the device. 
To quantify $\Gup/\Gp$, we employed a number of modeling assumptions. Firstly, 
the Fermi energy was fixed at $E_F=E_1/2$, corresponding to one half of the 
energy of the $n=1$ LL, the midpoint of the conductance plateau.
Secondly, a perfectly round inner contact and ballistic transmission 
were considered in order to estimate $\Gup = (4e^2/h) 2 j_m$, where $j_m \sim 
k_F R_{\textrm{in}}$ is the maximum total angular momentum \cite{yo_corbino}. 
Third, geometrical factors such as $R_{\textrm{in}}$ and $\Weff$ were scaled 
from our original device ($\alpha =1$). Finally, the value $T\approx0.9$ was 
extracted from the disordered conductance of \Fref{fig:fig2_te}c, from where 
we obtain $\mathcal{T} \approx 0.6$. 
\Fref{fig:fig2_te}d shows that, in pristine devices, $\Gup/\Gp$  decreases as 
$ \propto L^{-3/2}$. This decrease is due solely to $\Gup$ because, in a clean
device, $\Gp$ remains at the value $\approx 6(2e^2/h)$ for any $L$ as 
long as $E_F=E_1/2$. But, since the PMF scales inversely with $L$ for a fixed 
device shape, to keep this choice of $E_F$ requires its value to vary for 
different $L$. Hence, each $L$ in the figure corresponds to a different $k_F$, 
and these parameters vary inversely to each other; this causes the overall 
scaling shown by the black line in the figure. However, in the presence of the 
adsorbates, the 1D transport regime that occurs under pressure is expected 
to be much more sensitive to the disorder. The rapid decrease of the 
transmission probability $\mathcal{T}$ with $L$ in this case dominates over the 
variation of $\Gup$. As a result, the figure of merit, which is governed 
by the overall transmission probability through $\Gup/\Gp \propto 
j_m/\mathcal{T}$, increases with the channel length reaching values near 
12 for a device with $L = 400\unit{nm}$ ($R_{\textrm{in}} =  70\unit{nm}$, 
$\alpha =15$), a high ``on/off'' value considering the absence of a band gap in 
graphene.

\section{Conclusions}
In this work we set out to determine the optimal geometry of a graphene Corbino 
device that guarantees a PMF of nearly constant magnitude throughout most of 
the system when externally pressurized. Since the sign of the PMF has to 
inevitably alternate six times along any closed path containing the inner 
contact, the field cannot be strictly constant in the whole ring 
\cite{GuineaNP2010}. Yet, both its magnitude and absolute value can be made 
satisfactorily uniform. The first case is discussed in the \SI\ and 
we see that it is possible to require a specific sign and magnitude in most of 
the system, and the effect of the optimization process is to generate a geometry 
where the regions with the opposite PMF are much reduced (a more extreme spatial 
reduction is possible under higher pressures and higher target PMFs). If, on the 
other hand, the goal is only the PMF magnitude, but not its sign, the optimal 
geometries are the sixfold flower shapes such as in \Fref{fig:flower} that we 
have analyzed in detail. The spatial boundaries where the PMF changes sign for 
the latter are sharper and better defined, which leads to robust snake states 
and strictly one-dimensional transport over a range of pressures. This radially 
one-dimensional regime is signaled by the strict quantization of the conductance 
at $2e^2/h$ per channel that is seen to survive up to high values of $E_F$ (e.g. 
50--100\,meV, cf \Fref{fig:te_flor_str}), the characteristically 1D behavior of 
the DOS as a function of energy there, and corroborated by the real space 
profile of the LDOS.

The strain-induced perfectly conducting channels can be exploited in two 
different directions for electronic applications. On the one hand, they limit 
from below the current pinch-off effect and make these devices leaky because of 
their chiral nature. Our studies of the effect of disorder show that short-range 
defects can activate inter-valley scattering and reduce the residual conductance 
. Since their spatial location is set by the geometry, we can envisage this 
being done in a deliberate way through adsorbates, for example, so that only the 
regions where snake states develop under pressure become disordered. In this 
way, the conductance in the unpressurized state would remain unaffected, but the 
perfectly conducting channels would no longer exist under pressure which would 
significantly boost the on/off ratio.
The pressure sensitivity and its direct translation into current modulations, 
suggests its possible application in electromechanical sensing or transducers 
\cite{Zhou2013ElectrostaticGraphene}.
On the other hand, we have also seen an equally interesting perspective where 
snake-type states and their leaky residual conductance are not detrimental but, 
instead, functional: clean devices can be employed as sources of 
valley-polarized currents in graphene. Note that the Corbino ring 
(the device) is defined only by the indentation of the substrate, not by an 
actual patterning of graphene, and the inner and outer contacts at 
$R_{\textrm{in}}$ and 
$R_{\textrm{out}}$ are still the same sheet of graphene. Hence, the sheet can 
extend over 
large distances beyond the outer radius and these structures can act as local 
sources of valley-polarized currents for injection into the two dimensional 
graphene plane beyond $R_{\textrm{out}}$.

\begin{acknowledgement}
DAB acknowledges the support from Mackpesquisa and FAPESP under grant 
2012/50259-8, VMP that of the Singapore Ministry of Education Academic 
Research Fund Tier 2 under grant number MOE2015-T2-2-059, and AHCN the support 
of the National Research Foundation of Singapore under the Mid Size Centre 
Grant.
Numerical computations were carried out at the HPC facilities of the NUS Centre 
for Advanced 2D Materials.
\end{acknowledgement}

% --- For submission to the arXiv ----------------------------------------------
% \begin{suppinfo}
% 
% Details of the optimization procedure and equations; 
% optimized shapes for different values of dimensionless bending stiffness; 
% penalizing negative PMF; 
% annular Corbino geometries; 
% conductance of unpressurized devices; 
% snake states in real inhomogeneous magnetic fields; 
% conductance of optimized vs non-optimized Corbino devices; 
% scaling up devices and their transport characteristics; 
% disorder in the unpressurized flower device.
% 
% \end{suppinfo}
% ------------------------------------------------------------------------------

\bibliography{optimized_corbino}

% --- For submission to the arXiv ----------------------------------------------
% \newpage
% \section*{For Table of Contents Only}
% 
% \begin{figure*}
%   \centering
%   \includegraphics[width=0.6\textwidth]{Figs/Schematic.png}
% \end{figure*}
% ------------------------------------------------------------------------------

% ------------------------------------------------------------------------------
% SUPPORTING INFORMATION
% ------------------------------------------------------------------------------

\newpage


\section*{Supplementary material}

# Supporting Information

# Quantized transport, strain-induced perfectly conducting modes and valley filtering on shape-optimized graphene Corbino devices

Gareth W. Jones,[†] Dario Andres Bahamon,[‡] Antonio H. Castro Neto,[¶,§] and Vitor M. Pereira[*,¶,§]

[†]*School of Mathematics, The University of Manchester, Manchester, M13 9PL, England*
[‡]*MackGraphe – Graphene and Nano-Materials Research Center, Mackenzie Presbyterian University, Rua da Consolação 896, 01302-907, São Paulo, SP, Brazil*
[¶]*Department of Physics, National University of Singapore, 2 Science Drive 3, Singapore 117542*
[§]*Centre for Advanced 2D Materials, National University of Singapore, 6 Science Drive 2, Singapore 117546*

E-mail: vpereira@nus.edu.sg

Phone: +65 6601 3642

# Details of the optimization procedure and equations

In dimensionless terms (signified by an overbar, $\bar{\cdot}$), the full optimization problem to be solved is as follows:

$$\text{Minimize} \quad \bar{\mathcal{I}} = \frac{1}{\text{area}\,\bar{\Omega}} \iint_{\bar{\Omega}[c_1,\ldots,c_n]} (\bar{B}^2 - 1)^2 \, \mathrm{d}^2\bar{\boldsymbol{X}} + \eta \mathcal{I}^{\text{reg}}[c_1,\ldots,c_n], \tag{1}$$

subject to the six equations (valid for all admissable variations $\tilde{\cdot}$)

$$\frac{1}{\text{area}\,\bar{\Omega}} \iint_{\bar{\Omega}[c_1,\ldots,c_n]} \left[ \frac{\partial \tilde{v}_1}{\partial \bar{X}} \bar{N}_{11} + \frac{\partial \tilde{v}_1}{\partial \bar{Y}} \bar{N}_{12} \right] \mathrm{d}^2\bar{\boldsymbol{X}} = 0, \tag{2}$$

$$\frac{1}{\text{area}\,\bar{\Omega}} \iint_{\bar{\Omega}[c_1,\ldots,c_n]} \left[ \frac{\partial \tilde{v}_2}{\partial \bar{X}} \bar{N}_{12} + \frac{\partial \tilde{v}_2}{\partial \bar{Y}} \bar{N}_{22} \right] \mathrm{d}^2\bar{\boldsymbol{X}} = 0, \tag{3}$$

$$\begin{aligned}\frac{1}{\text{area}\,\bar{\Omega}} \iint_{\bar{\Omega}[c_1,\ldots,c_n]} &\left\{ \frac{\partial \tilde{w}}{\partial \bar{X}} \left( -\kappa \left( \frac{\partial \bar{M}_{11}}{\partial \bar{X}} + \frac{\partial \bar{M}_{12}}{\partial \bar{Y}} \right) + \bar{N}_{11} \frac{\partial \bar{w}}{\partial \bar{X}} + \bar{N}_{12} \frac{\partial \bar{w}}{\partial \bar{Y}} \right) \right. \\ &\left. + \frac{\partial \tilde{w}}{\partial \bar{Y}} \left( -\kappa \left( \frac{\partial \bar{M}_{12}}{\partial \bar{X}} + \frac{\partial \bar{M}_{22}}{\partial \bar{Y}} \right) + \bar{N}_{12} \frac{\partial \bar{w}}{\partial \bar{X}} + \bar{N}_{22} \frac{\partial \bar{w}}{\partial \bar{Y}} \right) + \tilde{w}\bar{p} \right\} \mathrm{d}^2\bar{\boldsymbol{X}} = 0, \end{aligned} \tag{4}$$

$$\frac{1}{\text{area}\,\bar{\Omega}} \iint_{\bar{\Omega}[c_1,\ldots,c_n]} \left[ \frac{1}{(1-\nu^2)} (\bar{M}_{11} - \nu \bar{M}_{22}) \tilde{M}_{11} + \frac{\partial \bar{w}}{\partial \bar{X}} \frac{\partial \tilde{M}_{11}}{\partial \bar{X}} \right] \mathrm{d}^2\bar{\boldsymbol{X}} = 0, \tag{5}$$

$$\frac{1}{\text{area}\,\bar{\Omega}} \iint_{\bar{\Omega}[c_1,\ldots,c_n]} \left[ \frac{1}{(1-\nu)} \bar{M}_{12} \tilde{M}_{12} + \frac{1}{2} \frac{\partial \bar{w}}{\partial \bar{X}} \frac{\partial \tilde{M}_{12}}{\partial \bar{Y}} + \frac{1}{2} \frac{\partial \bar{w}}{\partial \bar{Y}} \frac{\partial \tilde{M}_{12}}{\partial \bar{X}} \right] \mathrm{d}^2\bar{\boldsymbol{X}} = 0, \tag{6}$$

$$\frac{1}{\text{area}\,\bar{\Omega}} \iint_{\bar{\Omega}[c_1,\ldots,c_n]} \left[ \frac{1}{(1-\nu^2)} (\bar{M}_{22} - \nu \bar{M}_{11}) \tilde{M}_{22} + \frac{\partial \bar{w}}{\partial \bar{Y}} \frac{\partial \tilde{M}_{22}}{\partial \bar{Y}} \right] \mathrm{d}^2\bar{\boldsymbol{X}} = 0, \tag{7}$$

together with the additional definitions

$$\bar{N}_{11} = \bar{\varepsilon}_{11} + \nu\bar{\varepsilon}_{22}, \qquad \bar{N}_{12} = (1-\nu)\bar{\varepsilon}_{12}, \qquad \bar{N}_{22} = \nu\bar{\varepsilon}_{11} + \bar{\varepsilon}_{22}, \tag{8}$$

$$\bar{\varepsilon}_{11} = \frac{\partial \bar{v}_1}{\partial \bar{X}} + \frac{1}{2}\left(\frac{\partial \bar{w}}{\partial \bar{X}}\right)^2, \tag{9}$$

$$\bar{\varepsilon}_{12} = \frac{1}{2}\left(\frac{\partial \bar{v}_1}{\partial \bar{Y}} + \frac{\partial \bar{v}_2}{\partial \bar{X}} + \frac{\partial \bar{w}}{\partial \bar{X}}\frac{\partial \bar{w}}{\partial \bar{Y}}\right), \tag{10}$$

$$\bar{\varepsilon}_{22} = \frac{\partial \bar{v}_2}{\partial \bar{Y}} + \frac{1}{2}\left(\frac{\partial \bar{w}}{\partial \bar{Y}}\right)^2, \tag{11}$$

$$\bar{B}[\bar{\varepsilon}^{\mathrm{rec}}] = \frac{\partial}{\partial \bar{Y}}\left(\frac{\bar{\varepsilon}^{\mathrm{rec}}_{11} - \bar{\varepsilon}^{\mathrm{rec}}_{22}}{2}\right) + \frac{\partial \bar{\varepsilon}^{\mathrm{rec}}_{12}}{\partial \bar{X}}. \tag{12}$$

For a full derivation of this system in general, we refer the reader to reference 1, but we will give a brief overview here for convenience. $\boldsymbol{X} = (X, Y)$ are the Cartesian coordinates of the undeformed graphene flake. The two corresponding in-plane displacements are $v_1$ and $v_2$, and $w$ is the out-of-plane deflection. These, in turn, lead to the strain components $\varepsilon_{\alpha\beta}$ through (9)–(11), and in turn to the stress resultants $N_{\alpha\beta}$ using (8). This linear stress–strain relationship only holds for small values of strain, namely up to around 5–6%. The equations (2)–(7) are weak form representations of the standard Föppl–von Kármán relations for the deformation of an elastic plate. Though the typical statement of this law is in terms of two coupled fourth-order PDEs written in terms of a stress function, we do not use this representation here since the boundary conditions are displacement-based rather than force-based. Furthermore, the advantage of a weak-form formulation is that it is straightforward to discretize using finite element methods. Throughout this investigation we used linear finite elements on a triangular discretization of the material domain $\bar{\Omega}$; this means that the variables in question are parametrized by their values at the nodal points of the mesh, and their values in the interiors of the triangular elements are suitable linear interpolations of their values at the element nodes.[2]

In a standard bending plate theory, the bending moments of the plate at any point are assumed to be proportional to the plate curvature, which is in turn assumed to be the

second gradient of the out-of-plane displacement $w$. This is not achievable using linear finite elements, but we can keep the simplicity of this framework by rewriting the plate equations as a mixed variational principle, whereby the moment tensor components $M_{11}$, $M_{12}$, $M_{22}$ are assumed to be independent state variables in addition to the usual displacement variables (for full details, please see[1]). The equations are formulated to represent clamped boundary conditions, assuming that all three displacement components are zero on the boundary of $\bar{\Omega}$.

The leading-order term for the pseudomagnetic field $\bar{B}$ is given by a gradient of the strain tensor. However, the strain tensor defined by (9)–(11) is discontinuous if linear finite elements are used, and so its gradient would be undefined. Thus we must reconstruct a continuous strain field $\bar{\varepsilon}_{\alpha\beta}^{\text{rec}}$ from this discontinuous data, a process known as strain recovery (see[1,3]), and it is this which is differentiated to provide the PMF in (12).

The main goal of the method is to minimize the quantity $\bar{\mathcal{I}}$ in (1), which balances two effects: firstly we wish the square of the scaled PMF to be as close to unity as possible, which has the effect of penalizing shapes which produce large areas of near-zero PMF. The second term is a regularization term which penalizes intricate high-resolution oscillations in the outline shape (to which optimization techniques have a tendency to naturally approach), and produces smoother outlines. The regularization parameter $\eta$ tells us how strongly to weight the smoothness criterion. In our calculations we fixed this numerical constant to $10^{-5}$.

In short, the method demands that we vary the control variables $c_1, \ldots, c_n$ that define the shape, together with the state variables $\bar{v}_1$, $\bar{v}_2$, $\bar{w}$, $\bar{M}_{11}$, $\bar{M}_{12}$, $\bar{M}_{22}$ in such a way that the objective function $\bar{\mathcal{I}}$ is minimized subject to the elasticity constraints (2)–(7). The boundary conditions applied to the system are that the displacements are set to zero on nodes comprising the inner and outer boundaries. At the boundaries $\theta = \pm\pi/3$, we impose rotated periodicity conditions on these quantities (expressed in polar coordinates $R$, $\theta$), *i.e.*

$\bar{w}|_{\theta=\pi/3} = \bar{w}|_{\theta=-\pi/3}$, and similarly for $\bar{v}_R$, $\bar{v}_\theta$, $\bar{M}_{RR}$, $\bar{M}_{R\theta}$, $\bar{M}_{\theta\theta}$. This corresponds to

$$\bar{w}|_{\theta=\pi/3} = \bar{w}|_{\theta=-\pi/3}, \tag{13}$$

$$\begin{pmatrix} 1 & \sqrt{3} \\ -\sqrt{3} & 1 \end{pmatrix} \begin{pmatrix} \bar{v}_1 \\ \bar{v}_2 \end{pmatrix}\bigg|_{\theta=\pi/3} = \begin{pmatrix} 1 & -\sqrt{3} \\ \sqrt{3} & 1 \end{pmatrix} \begin{pmatrix} \bar{v}_1 \\ \bar{v}_2 \end{pmatrix}\bigg|_{\theta=-\pi/3}, \tag{14}$$

$$\begin{pmatrix} 1 & 2\sqrt{3} & 3 \\ -\sqrt{3} & -2 & \sqrt{3} \\ 3 & -2\sqrt{3} & 1 \end{pmatrix} \begin{pmatrix} \bar{M}_{11} \\ \bar{M}_{12} \\ \bar{M}_{22} \end{pmatrix}\bigg|_{\theta=\pi/3} = \begin{pmatrix} 1 & -2\sqrt{3} & 3 \\ \sqrt{3} & -2 & -\sqrt{3} \\ 3 & 2\sqrt{3} & 1 \end{pmatrix} \begin{pmatrix} \bar{M}_{11} \\ \bar{M}_{12} \\ \bar{M}_{22} \end{pmatrix}\bigg|_{\theta=-\pi/3}, \tag{15}$$

in terms of Cartesian components. The same conditions hold for the variations $\tilde{v}_1, \ldots, \tilde{M}_{22}$, so that (for example) the equations for the moment tensor at $\theta = \pi/3$ include relevant contributions from the boundary at $\theta = -\pi/3$.

To recover the physical values of the variables from their equivalent dimensionless quantities $\bar{\cdot}$, let $L$ be a typical length scale of the domain $\Omega$, $B_0$ to be the target PMF value, and $C$ and $D$ to be the stretching and bending moduli, respectively. We will invariably set $L$ to be the midpoint of the limits for the outer radius of the device, *i.e.* $L_{\text{out}}$ from equation (5) of the main text. Set

$$\varepsilon = \frac{aeB_0L}{\hbar c} \tag{16}$$

to be the typical scaling of the strain field, where $a = 1.42\,\text{Å}$ is the interatomic spacing of the graphene lattice, $e = 1.60 \times 10^{-19}\,\text{C}$ is the elementary charge, and $c \approx 3.37$ is a dimensionless parameter related to the rate of change in the electronic hopping in graphene.[1] The physical

quantities are then written in terms of their dimensionless values through

$$(X, Y) = L(\bar{X}, \bar{Y}), \qquad v_\alpha = L\varepsilon \bar{v}_\alpha, \qquad w = L\sqrt{\varepsilon}\bar{w}, \tag{17}$$

$$\varepsilon_{\alpha\beta} = \varepsilon\bar{\varepsilon}_{\alpha\beta}, \qquad \varepsilon_{\alpha\beta}^{\text{rec}} = \varepsilon\bar{\varepsilon}_{\alpha\beta}^{\text{rec}}, \qquad N_{\alpha\beta} = C\varepsilon\bar{N}_{\alpha\beta}, \qquad M_{\alpha\beta} = \frac{D\sqrt{\varepsilon}}{L}\bar{M}_{\alpha\beta}, \tag{18}$$

$$p = \frac{C\varepsilon^{3/2}}{L}\bar{p}, \qquad B = B_0 \bar{B}. \tag{19}$$

Note that, in dimensional terms, the objective function (1) becomes

$$\mathcal{I} = \frac{1}{\text{area } \Omega} \iint_{\Omega[c_1,\ldots,c_n]} (B[\boldsymbol{\varepsilon}^{\text{rec}}]^2 - B_0^2)^2 \, \mathrm{d}^2 \boldsymbol{X} + \eta \mathcal{I}^{\text{reg}}[c_1, \ldots, c_n], \tag{20}$$

confirming that $B_0$ does indeed correspond to the target PMF. Also, we note that we could, in principle, replace the PMF term in the objective function by a higher power such as $(B^2 - B_0^2)^4$, but this leads to a loss of sensitivity in the optimization procedure for values of $B$ near the target value $B_0$.

Other than Poisson's ratio $\nu$ and the dimensionless pressure $\bar{p}$, the only remaining physical parameter in the system is the dimensionless bending stiffness $\kappa$:

$$\kappa = \frac{D}{CL^2\varepsilon} = \frac{D\hbar c}{CL^3 aeB_0}. \tag{21}$$

The bending modulus $D$ of a hexagonal carbon lattice was calculated *ab initio* by Kudin et al.,[4] who found a value of $D = 1.46 \, \text{eV} = 2.34 \times 10^{-19} \, \text{N m}$ which will be used here. To calculate the stretching modulus $C = Eh/(1 - \nu^2)$, we use the results of Wei et al.,[5] who fitted a polynomial stress–strain relation to *ab initio* calculations up to strains of 50%. Their linear terms are, in our notation,

$$C = 358.1 \, \text{N m}^{-1}, \qquad C\nu = 60.4 \, \text{N m}^{-1} \qquad \Rightarrow \qquad \nu = 0.169. \tag{22}$$

Finally, note that though the method only provides the values of the graphene sheet's

displacement at the nodal values, it is trivial to calculate the piecewise linear interpolation of these nodal values to determine the displacement of an arbitrarily-positioned atom in the lattice. We work within the Cauchy-Born framework, where macroscopic displacements are directly mapped to all the atoms in the lattice (see, e.g., references 6,7 for a discussion of how to integrate corrections arising from non-zero displacements within the crystal unit cell).

## Optimized shapes for different values of dimensionless bending stiffness

As noted in the main text, and from eqs. (16), (19)$_1$ and (21), the optimized shapes depend only on the dimensionless quantities

$$\bar{p} = \frac{1}{C}\left(\frac{\hbar c}{ae}\right)^{3/2} \frac{p}{B_0^{3/2} L^{1/2}}, \qquad \kappa = \left(\frac{D}{C}\frac{\hbar c}{ae}\right) \frac{1}{L^3 B_0}. \tag{23}$$

In fact, if the bending stiffness $\kappa$ is varied while keeping $\bar{p}$ constant, there is very little variation in the shape. This is shown explicitly in Figure S1 that compares the optimal shapes obtained for different stiffness values spanning two orders of magnitude while keeping the other parameters as in the cases discussed in main text: $\bar{p} = 0.337$, $R_{\text{out}}$ allowed to vary in the range $261\,\text{Å} \pm 60\,\text{Å}$, and $R_{\text{in}}$ in the range $61\,\text{Å} \pm 15\,\text{Å}$.

## Penalizing negative PMF

It is interesting to note what happens if, instead of minimizing the integral of $(B^2 - B_0^2)^2$ over the domain $\Omega$, we choose an integrand $(B - B_0)^2$. With this expression, the solution must not only have the correct magnitude of PMF, but it must also penalize negative PMF values. Employing the same initial parameters used for the flower solution ($B_0 = 10\,\text{T}$, $p = 10^7\,\text{Pa}$, $L = 261\,\text{Å}$), we find that the solution is a triangular-shaped flake with an inverted interior

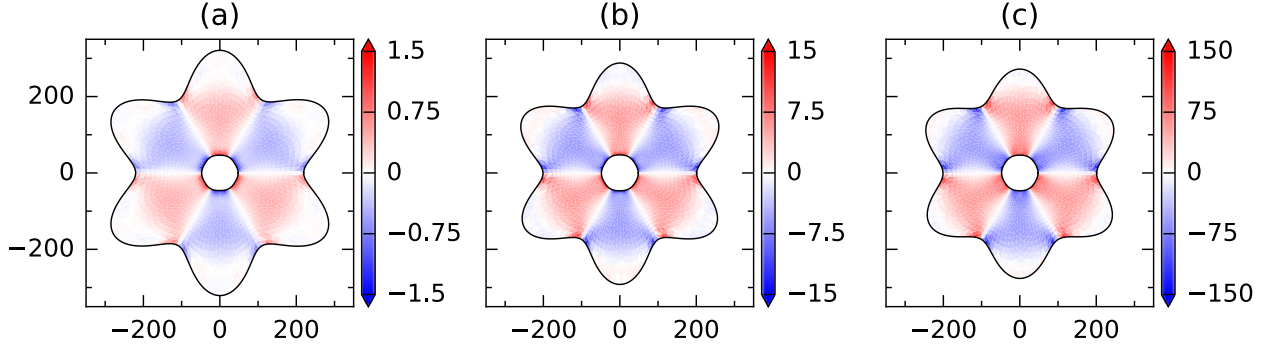

Figure S1: (color online) Output shapes at constant $\bar{p} = 0.337$, while $\kappa$ takes one of the three values $\kappa \in \{5.76 \times 10^{-4},\ 5.76 \times 10^{-5},\ 5.76 \times 10^{-6}\}$ in plots (a), (b), (c) respectively. Physical parameters used were, respectively, $p \in \{\sqrt{10}\,\text{bar},\ 100\,\text{bar},\ \sqrt{10}\,\text{kbar}\}$ and $B_0 \in \{1\,\text{T},\ 10\,\text{T},\ 100\,\text{T}\}$. Lengths are measured in Å and PMFs in T.

triangle, as displayed in Figure S2. It is clear that the optimization scheme has maximized the areas near $\theta \in \{\pi/2,\ -\pi/6,\ -5\pi/6\}$ where the PMFs are positive, and minimized the corresponding negative PMF regions around $\theta \in \{5\pi/6,\ \pi/6,\ -\pi/2\}$.

## Annular Corbino geometries

For comparison, we calculated the PMF associated with a circular annulus, *i.e.* $R_{\text{out}}(\theta) \equiv R_{\text{out}} = 1$ and $R_{\text{in}}(\theta) \equiv R_{\text{in}}$. In this case, the system of PDEs reduces to two ODEs by symmetry:

$$\kappa \left[ \bar{w}''''(\bar{R}) + \frac{2\bar{w}'''(\bar{R})}{\bar{R}} - \frac{\bar{w}''(\bar{R})}{\bar{R}^2} + \frac{\bar{w}'(\bar{R})}{\bar{R}^3} \right] - \bar{w}''(\bar{R}) \left[ \bar{v}'(\bar{R}) + \frac{(\bar{w}'(\bar{R}))^2}{2} + \frac{\nu \bar{v}(\bar{R})}{\bar{R}} \right] \\ - \frac{\bar{w}'(\bar{R})}{\bar{R}} \left[ \nu \bar{v}'(\bar{R}) + \frac{\nu(\bar{w}'(\bar{R}))^2}{2} + \frac{\bar{v}(\bar{R})}{\bar{R}} \right] + \bar{p} = 0, \quad (24)$$

$$\bar{v}''(\bar{R}) + \frac{\bar{v}'(\bar{R})}{\bar{R}} - \frac{\bar{v}(\bar{R})}{\bar{R}^2} + \bar{w}'(\bar{R})\bar{w}''(\bar{R}) + \frac{(1-\nu)(\bar{w}'(\bar{R}))^2}{2\bar{R}} = 0, \quad (25)$$

with $\bar{v} = \bar{w} = \bar{w}' = 0$ on $\bar{R} = R_{\text{in}}$ and $\bar{R} = 1$. Then the dimensionless PMF is given by

$$\bar{B} = \frac{\sin 3\theta}{2} \left[ -\frac{(3-\nu)(\bar{w}'(\bar{R}))^2}{2\bar{R}} - \frac{4\bar{v}'(\bar{R})}{\bar{R}} + \frac{4\bar{v}(\bar{R})}{\bar{R}^2} \right]. \quad (26)$$

S8

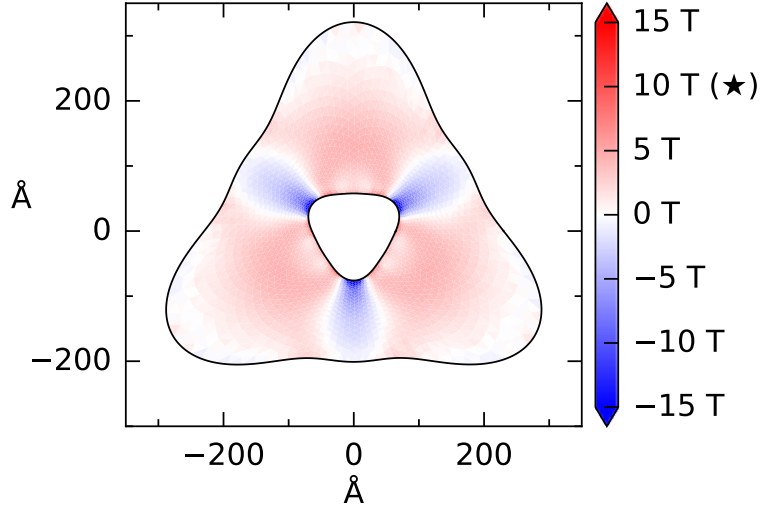

Figure S2: (color online) The PMF adopted by the graphene sheet when penalizing negative PMF values. The physical parameters were the same as those used in Fig. 2, and the target PMF was 10 T (indicated by ★ on the colorbar).

## Conductance of unpressurized devices

The conductance of the unpressurized flower is approximately proportional to the inner contact radius; this is clearly observed in Figure S3a where the conductance for differently-sized flower devices is plotted. The devices were generated from the optimized flower shape shown in Fig. 2 ($R_{\text{in}} \approx 47\,\text{Å}$ and $200\,\text{Å} \leq R_{\text{out}} \leq 322\,\text{Å}$), scaling inner and outer boundaries as $R_{\text{in}}^{\alpha}(\theta) = \alpha R_{\text{in}}(\theta)$ and $R_{\text{out}}^{\alpha}(\theta) = \alpha R_{\text{out}}(\theta)$. Larger $R_{\text{in}}^{\alpha}$ means a larger number of modes injected into the device; these modes are ballistically transmitted resulting in a higher conductance for larger $R_{\text{in}}^{\alpha}$. Given that the device is a completely open system, resonances do not appear; however, Fabry–Perot oscillations with periodicity $\Delta E = \pi \hbar v_F / W_{\text{eff}}$ arise indicating the presence of an effective width $W_{\text{eff}}$; this quantity is the effective channel length of our device.[8] For $\alpha = 1$ we found $W_{\text{eff}} = 223\,\text{Å}$ and $\Delta E \approx 0.03t$; for $\alpha = 2$ and $\alpha = 3$ we extracted $\Delta E^{\alpha=2} \approx 0.014$ and $\Delta E^{\alpha=3} \approx 0.01$, indicating that $W_{\text{eff}}^{\alpha} = \alpha W_{\text{eff}}$ as anticipated. Since the geometry is unaltered when the device is scaled up, the value of the conductance at the Dirac point does not change,[9] and $G(0) = 2.4(2e^2/h)$ for all values of $\alpha$. On the other hand, as observed in Corbino disks,[9,10] the conductance near the Dirac point is energy independent in an interval that shrinks as $R_{\text{in}}^{\alpha}$ increases, which is related to the nearly-circular geometry

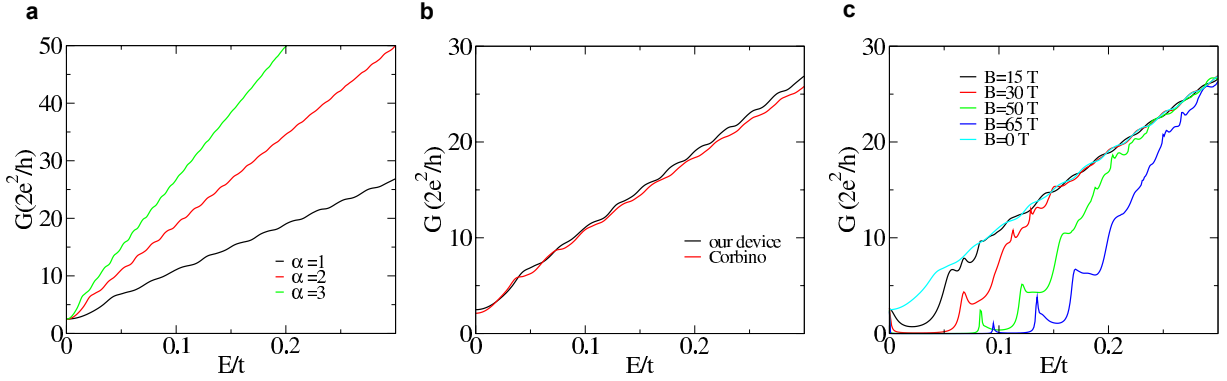

Figure S3: (color online) (a) Conductance of unpressurized devices scaled from the flower configuration (9)–(10) for scaling parameters $\alpha = 1, 2, 3$. (b) Comparison of the conductance of the flower device and a Corbino disk with $R_{\text{in}} = 47\,\text{Å}$ and $R_{\text{out}} = 261\,\text{Å}$, both unpressurized. (c) Conductance of the unpressurized flower device for different values of *real* magnetic fields constant in magnitude and sign. Note how the conductance under a homogeneous field soon reaches zero at low energies, as one expects in a Corbino geometry where no edge states contribute to the current.

of the inner contact. Based on the effect that $R_{\text{in}}^\alpha$ has on the conductance of the device, we compared in Figure S3b the conductance of our $\alpha = 1$ device that of an entirely circular Corbino disk of similar aspect ratio $R_{\text{in}}/R_{\text{out}}$ (specifically, $R_{\text{in}} = 47\,\text{Å}$ and $R_{\text{out}} = 261\,\text{Å}$ were chosen for the Corbino disk). The same behavior is observed in both devices, and their conductance very closely follow each other. This is explained by the fact that both are completely opened ballistic systems with circular or nearly-circular inner contact and similar inner/outer radius ratio.[9,11]

Switching on a (real for the moment) magnetic field $B$ introduces a new length scale in the problem. The resulting magneto-conductance can be understood by comparing the value of the cyclotron radius $r_c = \ell_B^2 k_F$ (where $\ell_B = \sqrt{\hbar/eB}$ is the magnetic length) with the variable distances between the inner and outer edges edges.[10,12,13] For $r_c$ smaller than half the minimum edge-to-edge distance ($r_c < (200-47)/2 = 76.5\,\text{Å}$), electrons emitted from the inner contact cannot reach the outer one and resonant tunneling through Landau levels (LL) at energies $E_n = (\hbar v_f/\ell_B)\sqrt{2n}$ emerges as the only transport mechanism. For example, in Figure S3c we see that at $B = 50\,\text{T}$ inner and outer edges are completely decoupled and

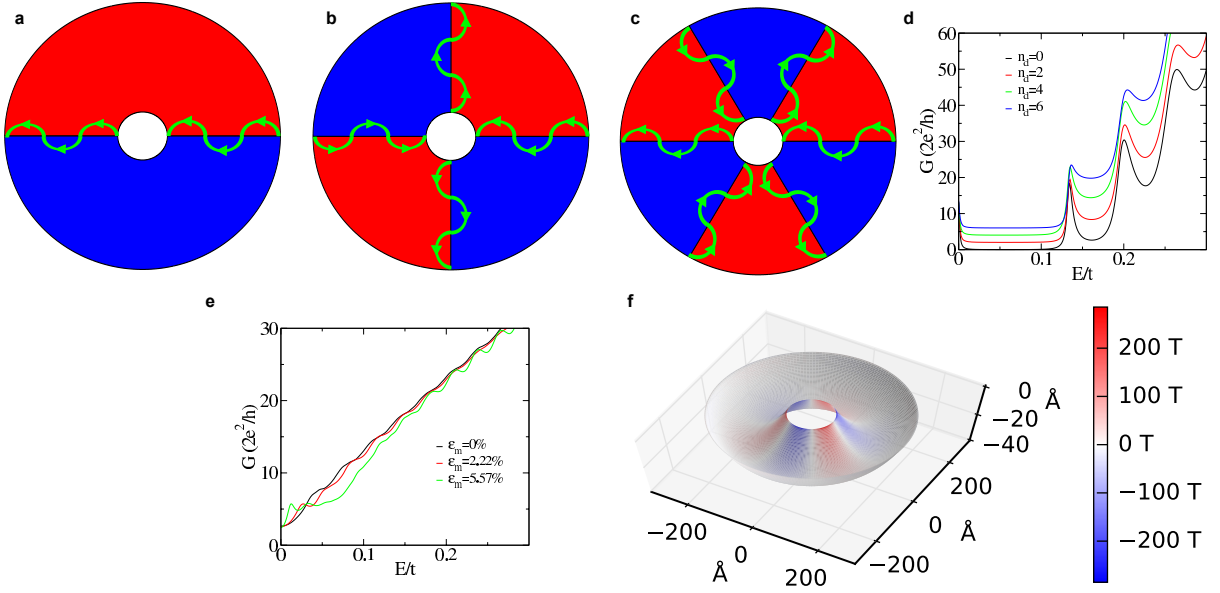

Figure S4: (color online) Schematic view of a Corbino disk and snake states in a real magnetic field with $n_d = 2$ (a), $n_d = 4$ (b), and $n_d = 6$ (c) sectors of alternating polarity (sign); red represents positive and blue negative. (d) Conductance of a perfectly annular Corbino disk ($R_{\text{in}} = 280\,\text{Å}$ and $R_{\text{out}} = 380\,\text{Å}$) under those inhomogeneous magnetic fields. Note that, in contrast with the case shown in Figure S3c for a uniform external field, the conductance is now perfectly quantized, proportionally to the number of polarity change boundaries. (e) Conductance of an annular Corbino disk ($R_{\text{in}} = 61\,\text{Å}$ and $R_{\text{out}} = 261\,\text{Å}$) under pressure (without any external magnetic field). (f) A surface plot (with exaggerated vertical scale) of an annulus (inner radius $61\,\text{Å}$, outer radius $261\,\text{Å}$) pressurized to $4\,\text{kbar}$, with a maximum vertical displacement of $20.8\,\text{Å}$, and a maximal strain of $5.76\%$ (colors indicate PMF magnitude).

resonant tunneling through the $n = 0/1$ LL is evident at the corresponding LL energies $E_0 = 0$ and $E_1 = 0.08t$ ($r_c \approx 53\,\text{Å}$).

## Snake states in real inhomogeneous magnetic fields

Previous studies of the transport properties of electrons under inhomogeneous magnetic fields have focused mostly on ribbon geometries.[14–17] Here, we calculate the conductance of a graphene Corbino disk of $R_{\text{in}} = 280\,\text{Å}$ and $R_{\text{out}} = 380\,\text{Å}$ in an inhomogeneous real magnetic field to identify the similarities and differences in the conductance of ideal, sharp field boundaries arising from a real magnetic field in relation to PMF. Note that, whereas

the spatial profiles of the real magnetic field discussed here are very unrealistic, those of the PMF discussed in the main text are, on the contrary, entirely realistic.

The field magnitude is set to $B_0$ over the entire annulus. The polarity or sign of $B_0$ alternates from positive to negative in neighboring sectors. This is schematically represented in Figures S4a–c for $n_d = 2$, $n_d = 4$ and $n_d = 6$ sectors respectively, identifying where the polarity was defined as positive (red) and negative (blue). Based on that color code we highlighted the snake states as well as their direction. The number of snake states pointing outwards is $N_s = n_d/2$, which has to be updated to $N_s = n_d$ when valley degeneracy is included. We set $B_0 = 125\,\text{T}$ in order to perfectly observe LL in the case that the field is homogeneous ($n_d = 0$). Figure S4c shows the obtained conductance. The entirely homogeneous case ($n_d = 0$) exhibits the anticipated resonant peaks at the LL energies. Increasing the number of magnetic field sectors leads to conductance plateaus at $n_d(2e^2/h)$ that are perfectly defined between the $n = 1$ and $n = 0$ LL.

## Conductance of optimized vs non-optimized Corbino devices

In order to illustrate that shape optimization is crucial to generate maximum confinement through PMF and to cause the emergence of snake states, we analyze here the transport characteristics of a non-optimized (strictly circular) Corbino device of equivalent dimensions, and under the same external pressure conditions. The pressurized Corbino disk of strictly circular geometry (radii $61\,\text{Å}$ and $261\,\text{Å}$) shown in Figure S4f has the conductance traces plotted in Figure S4e for different values of maximal strain $\varepsilon_\text{m}$. Even though at $\varepsilon_\text{m} = 5.57\%$ it displays a flattening of the conductance at low energies reminiscent of the behavior seen in the shape-optimized devices, this is explained by the much stronger PMF near the inner contact than within the annulus (see Figure S4f). These high PMF barriers on the inner contact force electrons to penetrate the device through regions where $B_s \approx 0$. But this

confinement is limited to the inner contact region and does not extend to the whole annulus: for example, LL are not formed in the central regions of the device and this flattening doesn't develop into a full and flat step upon increasing the device dimensions, as happens in the shape-optimized cases shown in Fig. 4(c) of the main text. As a result, we contend that the flower geometry possesses superior properties for potential applications.

## Scaling up devices and their transport characteristics

Of the many physical and geometric variables in the problem, only three are experimentally accessible. These are the device lengthscale $L$ (selected as $L_{\text{out}}$ from eq. (5) of the main text), the applied pressure $p$, and the target PMF $B_0$. In the main text we noted that these variables only affect the problem through the two dimensionless parameters $\bar{p}$ and $\kappa$ (provided in eq. (23) of this Supporting Information).

So, for instance, if one wished to find the applied pressure $p$ and target PMF $B_0$ for which a given lengthscale $L$ gave *exactly* the same flower shape through the optimization process, we would simply need to solve $\bar{p} = 0.337$ and $\kappa = 5.76 \times 10^{-5}$ for $p$ and $B_0$. In reality, however, this is too restrictive. Previously in this Supporting Information (Figure S1) we provided evidence that varying $\kappa$ has little effect on the optimal shape of the device. Indeed, the only effect of a larger $\kappa$ is that bending stiffness becomes a more important factor near the boundaries of the device. Since $\kappa = 5.76 \times 10^{-5}$ is small enough to confine bending effects to the boundary, the selected flower shape will still be close to optimal for any value of $\kappa$ smaller than this value, *i.e.*

$$\kappa \lesssim 5.76 \times 10^{-5}. \tag{27}$$

In the main text we show that the target PMF ($B_0$) is not strictly attained over the whole device; a smaller and nearly constant effective PMF ($B_{\text{eff}}$) appears instead. This effective field is responsible for the observed transport features, as we have established by determining $B_{\text{eff}}$ from the induced LL peaks in the LDOS or conductance traces. For our

flower geometry, our transport and DOS calculations show that the field extracted from the induced LL spectrum ($B_{\text{eff}}$) and $B_0$ are related by $B_{\text{eff}} \approx B_0/2.2$.

Now, let us fix $\bar{p} = 0.337$ in order to preserve the calculated flower shape. For a given lengthscale $L$ and applied pressure $p$, the formula

$$\bar{p} = \frac{1}{C}\left(\frac{\hbar c}{ae}\right)^{3/2} \frac{p}{B_0^{3/2} L^{1/2}} = 0.337 \qquad (28)$$

tells us what target PMF $B_0$ would have outputted the flower shape as optimal. Conversely, replacing $B_0 = 2.2 B_{\text{eff}}$ in (28), we can solve to find the effective PMF obtained by pressurizing the flower geometry of lengthscale $L$ with a given pressure $p$:

$$B_{\text{eff}} = \frac{1}{2.2 L^{1/3}}\left(\frac{\hbar c}{ae}\right)\left(\frac{p}{0.337 C}\right)^{2/3}. \qquad (29)$$

Substituting the solution for $B_0$ from (28) into the restriction (27), using (23), gives a restriction on $p$ for a given lengthscale $L$:

$$p \gtrsim \frac{0.337}{L^4 C^{1/2}}\left(\frac{D}{5.76 \times 10^{-5}}\right)^{3/2}. \qquad (30)$$

We wish to explore the parameter space $(L, p)$ for which snake states may be observed. First, electrons are to be injected from the inner contact with an energy $E \sim 0.5 E_1 = \frac{1}{2}\left(\frac{\hbar v_F}{\ell_B}\sqrt{2}\right)$. This guarantees that the energy of the electron is smaller than the energy of the first Landau level. We must thus ensure that Landau levels are formed in the bulk of the system, such that no transport occurs except through the radial interfaces where the PMF changes sign as snake-type states. Furthermore, in order to observe perfect quantization and perfect 1D-type transport, these modes must not be allowed to overlap.

Using a classical interpretation for the sake of obtaining specific estimates, these two criteria may be expressed as restrictions on the magnetic length $\ell_B$. Approximating the shape of the flower-shaped devices by an annulus of inner radius $\bar{R}_{\text{in}}$ and outer radius $\bar{R}_{\text{out}}$, the

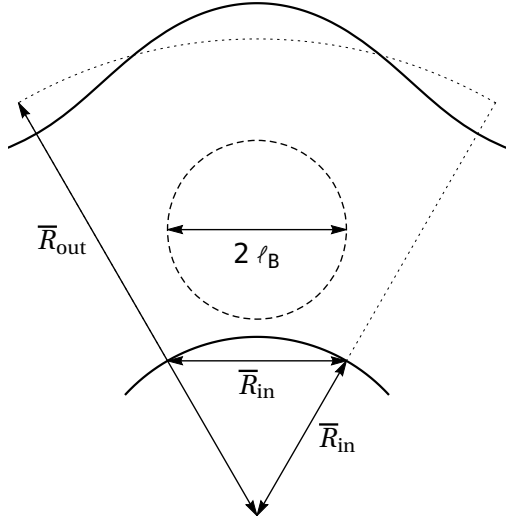

Figure S5: Schematic representation of a cyclotronic orbit in a petal (a 60° sector) of the flower device. One must have $\ell_B \lesssim \bar{R}_{\text{in}}/2$ in order to ensure that Landau levels are well developed within the inside regions of each petal, while simultaneously avoiding overlap between snake states propagating along neighboring radial paths.

diameter of the cyclotron orbit $2\ell_B$ must be shorter than both the annulus width $\bar{R}_{\text{out}} - \bar{R}_{\text{in}}$ and the inner width of a single $\pi/3$-sector (approximately $\bar{R}_{\text{in}}$). See Figure S5 for a graphical interpretation. Overbars here signify the mean radius of the shape.

For the dimensions of the optimal shape reported in the main text, $\bar{R}_{\text{in}} \sim 47\,\text{Å}$ and $\bar{R}_{\text{out}} \sim 245\,\text{Å} \sim 5\bar{R}_{\text{in}}$. Thus, for this particular shape, the restrictions on $\ell_B$ become

$$\ell_B < \frac{1}{2}\bar{R}_{\text{in}} \sim \frac{\bar{R}_{\text{out}}}{10}, \tag{31}$$

$$\ell_B < \frac{1}{2}(\bar{R}_{\text{out}} - \bar{R}_{\text{in}}) \sim \frac{4\bar{R}_{\text{out}}}{5}. \tag{32}$$

The first of these is the more restrictive condition, and we combine this with (29) and $\ell_B = \sqrt{\hbar/(eB_{\text{eff}})}$ to give

$$p \gtrsim \frac{0.337 C L^{1/2}}{(\bar{R}_{\text{out}}/10)^3}\left(\frac{2.2a}{c}\right)^{3/2}. \tag{33}$$

Since the preceding calculation uses a classical interpretation and geometry, it can only be considered as an order-of-magnitude estimate for the true limit as our particular problem is not in the classical regime. In order to more precisely fix this constraint, we take into

account the observation in Fig. 4a (main text) that the device with $\varepsilon_m = 6.11\%$ is at the threshold of perfect conductance quantization and strictly 1D transport (because the plateau is nearly perfectly developed at this pressurization already). We will demonstrate that, if the constraint (31) is relaxed slightly to $\ell_B < L/8$ (as $\bar{R}_{\text{out}} \sim L$), that particular device falls precisely on the threshold line and we, accordingly, use this latter condition to identify the range of parameters that we expect should lead to the same qualitative PMF and transport behavior in scaled devices. Under this assumption the constraint (33) becomes

$$p \gtrsim \frac{8^3 \times 0.337 C}{L^{5/2}} \left( \frac{2.2a}{c} \right)^{3/2}. \qquad (34)$$

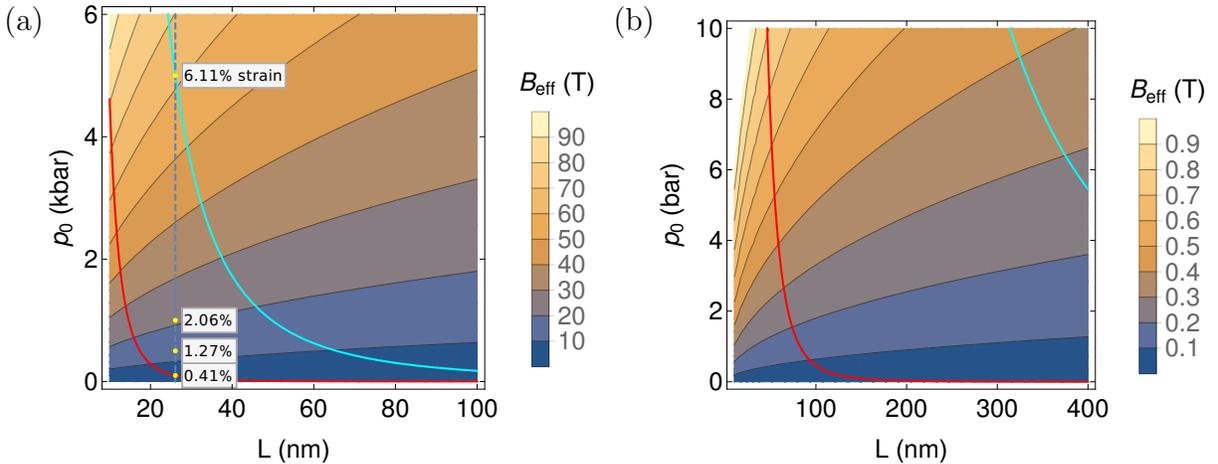

Figure S6: (a) A contour plot of the effective PMF $B_{\text{eff}}$ for a range of small lengthscales $L$ and applied pressures $p$ up to 6 kbar. The constraints (30) and (34) on $p$ are superimposed as red and cyan curves respectively; these constraints are satisfied in the region of parameters above each line. The four highlighted configurations correspond to the configuration optimized for $\varepsilon_m = 0.41\%$ and the three strained configurations investigated in Fig. 4(a)–(b) of the main text, all at $L = L_{\text{out}} = 26.1$ nm. (b) As plot (a), but for a larger (and more experimentally realizable) range of lengthscales $L$ (note the different units of the vertical axes).

In Figure S6(a) we plot the effective PMF from eq. (29) for a range of small devices (lengthscales up to 100 nm). On this we superimpose the two curves corresponding to the restrictions on $p$, from relations (30) and (34). Values of $p$ above the former (red) curve have small enough bending stiffness effects that the originally calculated device will still have the optimal shape; this constraint is easily satisfied for essentially all experimentally

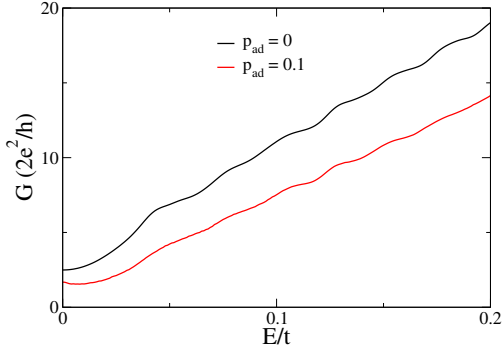

Figure S7: Conductance of the unpressurized device with and without adatoms.

relevant lengthscales and pressures. Values of $p$ above the cyan curve are needed in order to observe strictly one-dimensional transport. Note that the $\varepsilon_\mathrm{m} = 6.11\%$ solution lies directly on the cyan curve, confirming our previous statement regarding the transport characteristics of this device reported in Fig. 4a of the main text. In Figure S6(b) we have plotted the same results but for a larger range of lengthscales (up to $400\,\mathrm{nm}$). A device of lengthscale $L = 400\,\mathrm{nm}$ has an inner radius of $78\,\mathrm{nm}$, resulting in a pattern which is perfectly feasible with standard lithography. At this lengthscale, the minimum pressure for the observation of one-dimensional transport is found at the intersection with the cyan curve, *i.e.* at $p = 5.45\,\mathrm{bar}$. This results in an effective PMF of $0.26\,\mathrm{T}$ and (by design) a magnetic length of $\ell_B = 50\,\mathrm{nm}$.

Finally, note that, as follows from this discussion, it is not the magnitude of the generated PMF alone that determines whether the 1D transport regime is realized or not. It is rather the combination of the field magnitude, through its influence in the magnetic length and "orbit" size, with the characteristic device dimensions that determines the favorable conditions for the development of robust LL in the bulk of the device, and decoupled snake-type states along the radial interfaces.

## Disorder in the unpressurized flower device

In Figure S7 we show the conductance of the $\alpha = 1$ unpressurized flower device in the presence of adatoms, the adatoms are distributed over the whole area of the device. We can see that the conductance is reduced by $\sim 60\%$ for $E_F = 0.04t$, . For larger devices this value

should be increased given that the number of angular momentum channels $j_m \propto R_{\text{in}}$ and the conductance is proportional to $j_m$.

%
\AtBeginShipout\AtBeginShipoutDiscard

\end{document}